\documentclass[a4paper,twocolumn,11pt,accepted=2023-07-09]{quantumarticle}
\pdfoutput=1
\usepackage[utf8]{inputenc}
\usepackage[english]{babel}
\usepackage[T1]{fontenc}
\usepackage{amsmath}
\usepackage{hyperref}

\usepackage[a4paper,top=3cm,bottom=2cm,left=3cm,right=3cm,marginparwidth=1.75cm]{geometry}

\input{Qcircuit}

\usepackage{amsmath}
\usepackage{amsfonts}
\usepackage{graphicx}
\usepackage[colorinlistoftodos]{todonotes}

\usepackage{authblk}  

\usepackage[normalem]{ulem}

\usepackage{mathtools}

\usepackage{amsthm}  

\newtheorem{criterion}{Design criteria}
\newtheorem{algo}{Algorithm Sketch}

\usepackage{enumitem}  

\usepackage{tikz}
\usetikzlibrary{decorations}
\usetikzlibrary{decorations.pathmorphing}
\usetikzlibrary{decorations.pathreplacing}
\usetikzlibrary{decorations.shapes}

\usepackage{forloop}
\usepackage{scalerel}
\usetikzlibrary{backgrounds}
\usetikzlibrary{shapes.misc}

\DeclarePairedDelimiter{\ceil}{\lceil}{\rceil}

\newcommand{\la}{\langle}
\newcommand{\ra}{\rangle}
\newcommand{\kb}{\ra\la}
\newcommand{\sx}{X}
\newcommand{\sy}{Y}
\newcommand{\sz}{Z}
\newcommand{\pihalf}{\frac{\pi}{2}}

\newcommand{\fint}{{\breve{f}}}

\newcommand{\m}{\mathcal}
\newcommand{\mA}{\mathcal A}

\newcommand{\mP}{\mathcal P}

\newcommand{\eq}{\mathrm{EQ}}

\renewcommand{\O}{\Omega}

\newcommand{\tranasymm}[3]{\m T^{(#1 \leftarrow #2)}_{#3}} 
\newcommand{\tran}[3]{\m T_{#3}^{(#1 \leftrightarrow #2)}} 
\newcommand{\supr}[1]{^{(#1)}}

\newcommand{\tr}{\textrm}

\renewcommand{\a}{\alpha}
\renewcommand{\b}{\beta}

\title{Encoding trade-offs and design toolkits in quantum algorithms for discrete optimization: coloring, routing, scheduling, and other problems}
\author[1]{Nicolas PD Sawaya$^*$}
\author[2]{Albert T Schmitz}
\author[3,4]{Stuart Hadfield}
\affil[1]{Intel Labs, Intel Corporation, Santa Clara, California 95054, USA [\textbf{$^*$nicolas.sawaya@intel.com}]}
\affil[2]{Intel Labs, Intel Corporation, Hillsboro, Oregon 97124, USA}
\affil[3]{Quantum Artificial Intelligence Laboratory, 
NASA Ames Research Center, Moffett Field, California 94035, USA}
\affil[4]{USRA Research Institute for Advanced Computer Science, 
Mountain View, California, 94043, USA}

\begin{document}
\maketitle

\begin{abstract}
Challenging combinatorial optimization problems are ubiquitous in science and engineering. Several quantum methods for optimization have recently been developed, in different settings including both exact and approximate solvers. 
Addressing this field of research, this manuscript has three distinct purposes. First, we present an intuitive method for synthesizing and analyzing discrete (\textit{i.e.,} integer-based) optimization problems, wherein the problem and corresponding 
algorithmic primitives are 
expressed using a discrete quantum intermediate representation (DQIR) that is encoding-independent. This 
compact 
representation often allows for more efficient problem compilation, automated analyses of different encoding choices, easier interpretability, more complex runtime procedures, and richer programmability, as compared to previous approaches, which we demonstrate with a number of examples. 
Second, we perform numerical studies comparing several qubit encodings; the results exhibit a number of preliminary trends that help guide the choice of encoding for a particular set of hardware and a particular problem and algorithm. 
Our study includes problems related to graph coloring, the traveling salesperson problem, factory/machine scheduling, financial portfolio rebalancing, and integer linear programming.  
Third, we design low-depth graph-derived partial mixers (GDPMs) up to 16-level quantum variables, demonstrating that compact (binary) encodings are more amenable to QAOA than previously understood. 
We expect this toolkit of programming abstractions and low-level building blocks to aid in designing quantum algorithms for discrete combinatorial problems.

\end{abstract}

 \tableofcontents

\section{Introduction}\label{sec:intro}

Combinatorial optimization problems are ubiquitous across science, engineering, and operations research, encompassing diverse problem areas such as scheduling, routing, and network analysis, among others \cite{papadimitriou1998combinatorial}. 
This has led to 
much interest in the potential for quantum advantage for hard optimization tasks, in different settings including exact, approximate, and heuristic solvers \cite{grover1996,hogg2000quantum,farhi2014quantum,hastings2018exact,albash2018adiabatic,hadfield19_qaoa,hauke2020perspectives}.
The past few years in particular have seen 
development of novel quantum approaches for 
tackling these problems, with much focus on constraint satisfaction problems over 
binary variables, such as the commonly studied MaxCut problem \cite{farhi2014quantum}.  
However, from the application perspective,
a wide variety of important optimization problems 
are more naturally expressed over sets of discrete (typically integer) variables. This can add an additional layer of complexity when 
applying 
and implementing existing quantum algorithms, partly because there are many ways to encode a discrete variable into 
qubits, qudits, or other hardware, with different resource and performance tradeoffs.

Indeed, as a practitioner 
may have many algorithmic choices---regarding for instance the encoding, algorithm class, and 
parameters---it is vital to develop conceptual tools and software approaches that help prepare and implement algorithms for discrete optimization problems on quantum computers. Such tools can be useful for automating analyses of different approaches (such as different encodings), but they can also lead to superior programmability, which in turn may yield more efficient compilation and runtime implementations. 
In particular, seeking cleaner separations of programming layers is important towards enabling a broad community of practitioners who may not be experts in quantum mechanics or other low-level details \cite{Svore2006_layered,ittah21_dataflow,Shaydulin21_qaoakit}.

\begin{figure}
    \centering
    \includegraphics[width=0.35\textwidth]{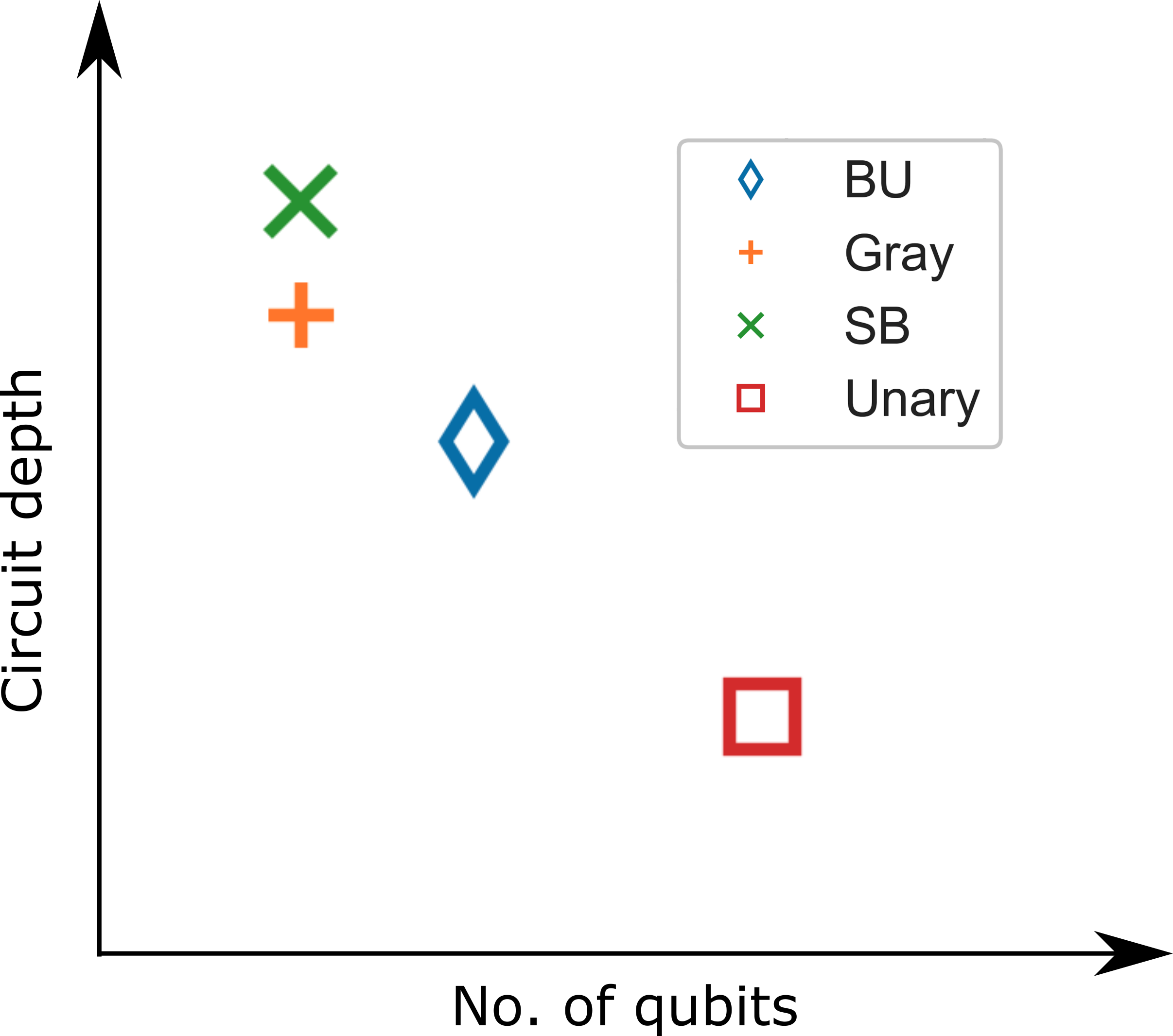}
    \caption{Illustration of a hypothetical Pareto front showing the resource trade-offs inherent in the encoding choice for a quantum algorithm. The variation between encodings can be very large with respect to both depth and space. For example, unary codes require $d$ qubits while
    compact (Gray and standard binary) codes use only $\log_2 d$ qubits, for encoding a discrete variable of cardinality $d$. However, the unary encoding tends to require shorter depths. Block unary (BU) encodings attempt to interpolate between unary and compact. These trade-offs between space and depth mean that different encodings are appropriate for different quantum hardware instances. As we demonstrate empirically, real problem instances show many exceptions to the trend of this image; for example, compact codes are sometimes superior in both qubit count and circuit depth.
    }
    \label{fig:pareto}
\end{figure}

Regarding encoding choice, we note that one may encode a variable into qubits in many different ways, each of which may have favorable properties for different hardware. For example, one encoding may be 
advantageous for a many-qubit device with lower available circuit depth, while another may be 
preferable 
for a 
device with more available depth but fewer qubits (see Figure \ref{fig:pareto}). Therefore it is useful to have a framework that can be used to automate the mapping, compilation, and analysis of a given encoding choice.

\begin{figure*}
    \centering
    \includegraphics[width=.8\textwidth]{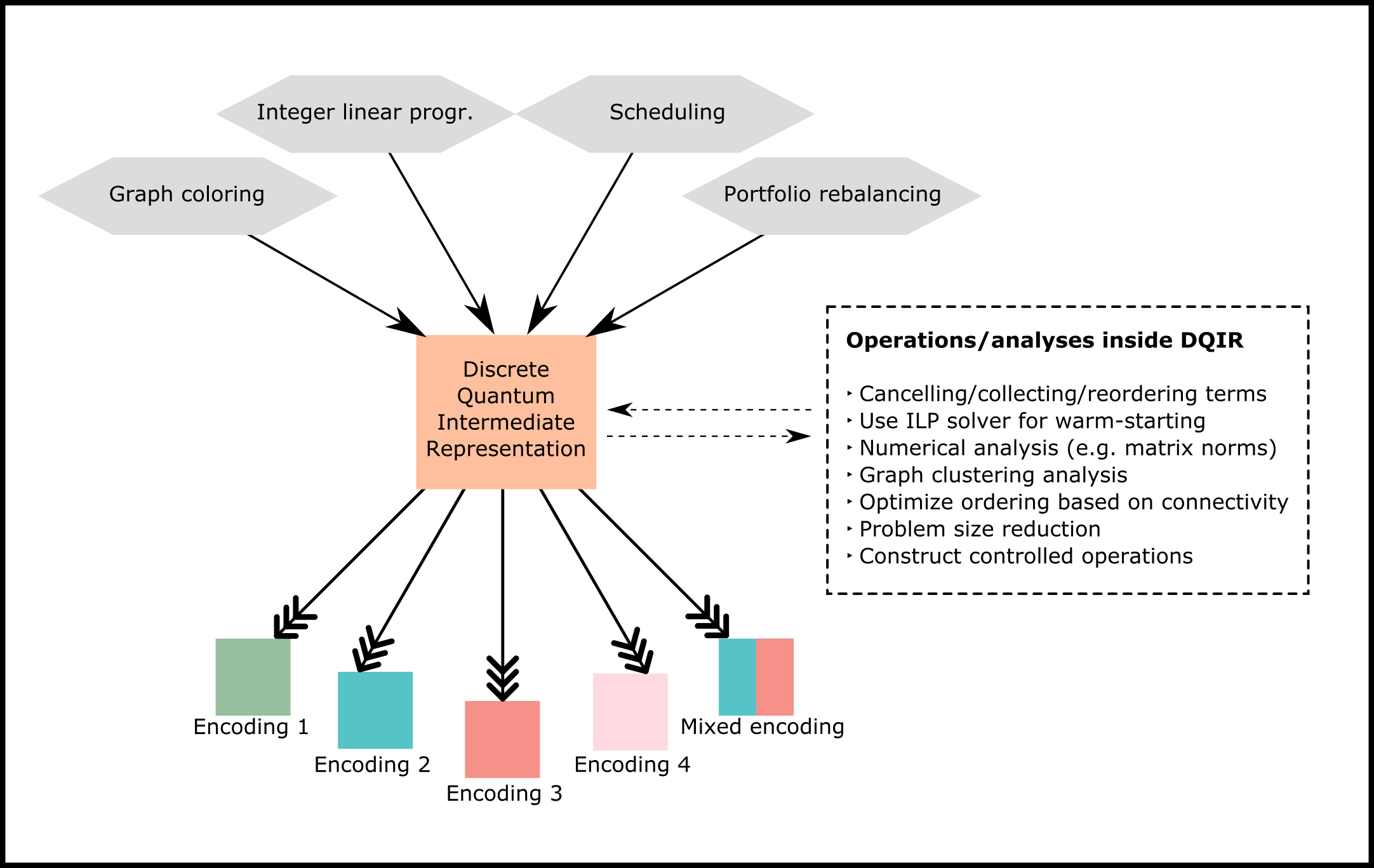}
    \caption{Schematic of the use of DQIR when preparing discrete combinatorial problems for solving on a quantum computer. Discrete optimization problems may be mapped to DQIR, where several operations may take place. The mixed encoding refers to having different encodings for different quantum variables.}
    \label{fig:prob-ir-enc}
\end{figure*}

In the current work we (a) introduce an intuitive and efficient framework (an \textit{intermediate representation}) for constructing and implementing quantum algorithms for discrete optimization problems, 
including generalizing a number of existing results from the Boolean cube to more general discrete domains;   
(b) provide a pedagogical resource including an informal dictionary of useful 
primitives and relations in this general setting; (c) numerically analyze which encodings are advantageous in which scenarios; and (d) present what are to our knowledge the first ultra-low-depth designs of QAOA mixers for standard binary and Gray encodings 
of integer variables.
We will demonstrate how our framework provides a 
compact and practically useful representation of these problems. 
Figure \ref{fig:prob-ir-enc} gives a schematic of the workflow of our \textit{discrete quantum intermediate representation} (DQIR). DQIR is useful for preparing, manipulating, and analyzing problem instances independently of hardware implementations, while also automating the conversion to and analysis of encodings for the purpose of choosing the 
most advantageous one (e.g., given the resource constraints of a fixed real-world device).

Our work builds off of 
and extends
a number of previous works. In particular, \cite{sawaya20_dlev} which  studied encoding procedures and pitfalls as well as compilation tradeoffs
for $d$-level systems in the context of quantum simulation, \cite{hadfield18_repres} which formally defined basic primitives and Hamiltonian mappings for the binary optimization case, \cite{hadfield19_qaoa} that 
studied the design of quantum approaches for discrete optimization, including the one-hot and standard binary mappings for a diverse set of standard problems. While intermediate representations have been introduced for many aspects of quantum compilation \cite{Hietala19_ir,Nguyen21_mlir,McCaskey21_mlirdialect,Svore2006_layered,cross17_openqasm,ittah21_dataflow}, DQIR is intended for use specifically for problems defined over a domain of discrete variables.
Here we attempt to unify and extend these viewpoints into a more general but more user-friendly framework. We then demonstrate how DQIR facilitates more efficient compilation and analysis over previous approaches. Some of the 
constructions and encodings presented are novel in the context of quantum optimization. 
A further 
contribution is the comparison of circuit depths for several encodings over a range of standard problems and subroutines for commonly occurring domains. To our knowledge such a systematic numerical analysis 
has not been published previously, and we anticipate the results to be  directly useful to practitioners in the field.

It is useful to note some technical differences between physics simulation of $d$-level particles \cite{sawaya20_dlev,sawaya20_connect} (phonons \cite{macridin18a,mcardle19_qvibr,ollitrault20_reiher_qvibr,sawaya2021ir}, photons \cite{kottmann2021qcadoptics}, spin-$s$ particles \cite{lora2016spins}, etc.) and discrete combinatorial problems. In physics simulations the Hamiltonian itself usually contains non-diagonal operators relative to the computational basis, for example bosonic creation and annihilation operators. These non-diagonal operators often make the largest contribution to resource requirements \cite{sawaya20_dlev}. On the other hand, for classical combinatorial 
problems the cost function 
is typically 
mapped to a \textit{diagonal} operator, and 
there is often significant flexibility in choosing 
non-diagonal operators suitable for 
realizing a given quantum algorithm. 
This flexibility makes it easier 
to reduce the resource requirements in the optimization setting. On a related one, the measurement problem \cite{mcclean2016theory,verteletskyi2020measurement,cerezo2021variational,fedorov2022vqerev}---namely, that many measurements 
in many different bases 
are required to determine $\bra{\psi}H\ket{\psi}$ in physics simulation---is not nearly as much of a bottleneck in classical optimization problems for which the cost function is a diagonal operator 

There have been 
numerous studies on quantum approaches for particular discrete optimization problems \cite{lucas2014ising,hadfield19_qaoa}, including for problems related to graph coloring 
\cite{Oh19_multicoloring,wang20_xymixers,Tabi20_color,fuchs21_maxkcut}, planning \cite{OGorman16_planning,Stollenwerk20_planning,stollenwerk20_airtraffic}, scheduling \cite{crispin13_vehicle,venturelli15_jobshop,tran16_schedaaai,Krzysztof20_railway,Dalyac21_EVs,amaro21_jobshop,plewa2021variational}, routing \cite{glos20_space_eff,Salehi21_tsp}, integer programming \cite{Bernal20_ip}, and option pricing \cite{hodson19_portf, Ramos21_unaryoption}. Though most of these works employ either a unary-style or more compact binary-style qubit encoding of the problem variables, some have considered multiple encodings in the same work \cite{hadfield19_qaoa,Tabi20_color,glos20_space_eff,tamura21_knapsack,botelho21_midcirc}.
Unlike 
most previous studies, our representation and methods are presented at a higher layer that is encoding-independent; therefore, one can in principle reuse mappings and primitives across different quantum hardware where different encodings or algorithms may be most suitable, as well as in some cases across different problems.  
Indeed, a further advantage is that the lower layers are not restricted to be physical-qubit-based, and DQIR can easily envelop qudits (i.e., $d$-level quantum system, where $d$ may be not necessarily match the problem domain size), quantum fault-tolerance (i.e., encoded logical qubits), as well as continuous variable quantum computers or other more exotic proposals.

Critically, despite considerable effort it remains unclear exactly under which circumstances or for which problems or to what degree one may possibly achieve quantum advantage for combinatorial optimization, in the  near-term and beyond \cite{wang2018quantum,hadfield2018quantum,hadfield19_qaoa,hastings2019classical,bravyi2020obstacles,dalzell2020many,franca2020limitations,zhou2020quantum,barak2021classical,bittel2021training,marwaha2021bounds}.  
We do not 
try to 
tackle the challenging questions related to performance in this work, and  
focus instead on mathematical tools that may be practically useful toward algorithm design and implementation, especially as more sophisticated and diverse quantum hardware platforms become available in the coming years.  




This paper is organized as follows. In Section \ref{sec:dqir} we define the primitives of the discrete quantum intermediate representation, discuss encodings and 
important subroutines, and summarize procedures for mapping the problem to hardware-specific (especially qubit-based) representations. In Section~\ref{sec:algo_compon} we overview several standard quantum approaches to optimization and consider their various required components, including 
mixers, penalties, and choice of initial state, while discussing best practices in each aspect. In Section \ref{sec:gdm} we introduce the novel concept of graph-derived partial mixers (GDPMs) in order to design specific 
resource-friendly mixers 
applicable to various problem classes. In Section \ref{sec:problems} we express five general classes of discrete optimization problem in terms of DQIR, highlighting the compactness and intuitiveness of the resulting expressions. In Section \ref{sec:results} we then present numerical results for some of these problems for 
implementing several common operators derived 
in multiple encodings, and discuss the various resulting resource trade-offs. Finally, in section \ref{sec:concl} we 
elaborate on the utility of our approach and 
discuss several future directions such as the incorporation of noise and hardware topology into our framework.

\section{Discrete quantum intermediate representation}\label{sec:dqir}

Here we formally introduce DQIR for quantum optimization algorithms, and beyond. 
There are several appealing reasons to use a DQIR in a compilation workflow. First, it provides a path to automated methods for encoding a range of problem types into any user-defined encoding. Instead of deriving conversions to qubit operators for each new encoding, as has been done in most previous work, one may implement any new encoding simply by defining a new integer-to-bit function. Second, a hardware-agnostic representation helps facilitate the interfacing with new devices, for example hardware with a novel topology or non-standard devices that use qutrits, ququads, or higher-order qudits \cite{ozguler2022numerical,deller2022quditqaoa}. Third, it 
can be more efficient to perform algebraic manipulations inside DQIR, 
because often the resulting terms are simpler and fewer. Finally, several problem analyses and preparation steps are more conceptually natural and can be calculated with fewer operations in DQIR, as we demonstrate with the examples considered in Section~\ref{sec:problems}.

\subsection{Discrete functions and optimization problems}
\label{sec:dqirProblemDefn}

We consider real-valued functions 
%
\begin{equation}
\begin{split}
f(\mathbf x) \equiv f(x_1,x_2,\dots, x_M),\\
\end{split}
\end{equation}
where we sometimes use 
 $\fint(\mathbf x)$ to denote the special case of functions taking values in $\{0,1,.\dots,k-1\}\equiv \mathbb Z_{k} $, 
over a domain of discrete variables, $x_\a \in D_\a$,  
\begin{equation}
\mathbf D \equiv D_1 \times \dots \times D_M.
\end{equation}
%
Such domains are isomorphic as sets to subsets of 
integers
\begin{equation}
\mathbf D \cong \mathbb Z_{d_1} \times \dots \times \mathbb Z_{d_M}
\end{equation}
%
so for simplicity 
we will assume 
integer variable domains for most of this work. 
Note that the domain cardinalities are often (but not always) independent of the number of problem variables; the familiar setting of combinatorial optimization over binary variables corresponds to the case $d_\alpha=2$. Similarly, the important special case of Boolean functions  $\fint(\mathbf x)\in\{0,1\}$ corresponds to the case $k=2$.  


\paragraph{Problem cost functions and constrained optimization.}
For a combinatorial optimization problem, we are typically given some representation of a function $f$ we seek to extremize as part of the problem input. For example, we may be given a set of clauses, functions $\{f_i\}$ 
each acting on a subset of the variables, from which $f$ is constructed using a suitable operation on the target space, such as $\fint =\sum_i \fint_i$ in a constraint satisfaction problem, or $\fint=\bigwedge_i \fint_i$ for (Boolean) satisfiability. Generally we say 
a family of functions 
in a given representation  
is \textit{efficiently represented} (as input) if it uses a number of variables that is polynomially scaling in the 
number of bits required to describe elements of the domain (in which case the usual notions of algorithmic efficiency apply). 

Additionally, $f(x_1,x_2,\dots)$ may be subject to a set of 
\textit{hard constraints}, such as equality constraints
\begin{equation}\label{eq:equal_constr}
\{g_i(x_1,x_2,\dots) = 0\}
\end{equation}
and/or inequality constraints 
\begin{equation}\label{eq:inequal_constr}
\{h_l(x_1,x_2,\dots) \geq 0\},
\end{equation}
which must be satisfied by any 
potential solution. 
%
Hard constraints such as \eqref{eq:equal_constr} and \eqref{eq:inequal_constr} hence induce 
a \textbf{feasible subspace} 
of the original 
domain (and corresponding Hilbert space), 
%
%
\begin{equation}
\mathbf D_{\text{feas}} \equiv \{\mathbf x_{\rm{feasible}}\}  \subseteq \mathbf{D},
\end{equation}
which may depend on the particular problem instance. 
Generally, hard constraints may be given as part of the problem input, 
or may additionally arise as various problem encoding choices are made. 
Note that while in principle hard constraints may be absorbed into a (possibly complicated) redefinition of the underlying domain of $f$, 
it is often advantageous to define simpler domains that do not depend on the particular instance and treat hard constraints 
via algorithmic primitives such as penalty terms or constraint-preserving mixers, as we discuss in Section \ref{sec:algo_compon}.

For the optimization setting our goal is to minimize $f(\mathbf x)$ over the feasible subspace, \textit{i.e.} subject to the hard constraints. (The maximization case is similar.) 
For a variety of important applications, 
in particular 
NP-hard optimization problems, it is believed that neither classical nor quantum computers can efficiently solve these problems optimally, for arbitrary problem instances. In such cases we may employ algorithms with super-polynomially scaling resources, or settle for efficiently obtained approximate solutions, where 
the goal is to find a configuration with function value as low as possible. For the latter, we may employ approximation algorithms, where a guarantee to achieved solution quality is known, or heuristics, where such a guarantee may not be; see for example \cite{hadfield2017quantum} for a more detailed discussion of quantum heuristics for 
approximate optimization. 

We highlight here one important example problem class. 
A variety of industrially important problems are expressible as what we call \textit{permutation problems}, for which 
%
we seek to optimize a function over possible permutations $\tau_k$, which we represent 
with strings of integers as 
\begin{equation}\label{eq:x_permut}
\mathbf x_{\tau_k} = (\tau_k(1),\tau_k(2),\cdots,\tau_k(M)).
\end{equation}
%
We define the 
space of 
permutations as
\begin{equation} \label{eq:perm_space}
\{\mathbf x\}_{\rm{perm}} = \{ \mathbf x_{\tau_k} | \tau_k \in S_M \}
\end{equation}
where $S_M$ is the permutation group on $M$ objects. 
In this example, 
infeasible
strings are those in which any integer appears twice. (For our purposes it is not necessary to consider the many forms of constraint $g_k$ that might be used to induce this feasible subspace of permutations.)
A subset of the problems considered in this work are permutation problems, namely scheduling and routing problems, which may come with additional feasibility constraints in practice. 

\paragraph{Hamiltonians representing functions.}
Following \cite{hadfield18_repres} we say a Hamiltonian $H_f$ represents a real function $f()$ on $\mathbf D$ if it acts diagonally 
\begin{equation} \label{eq:fcnRep}
 H_f\ket{\mathbf x} = f(\mathbf x)\ket{\mathbf x} = f(x_1,\dots,x_m)\ket{x_1}\otimes\dots\otimes\ket{x_m}
 \end{equation}
 for every 
basis state $\ket{\mathbf x}$, $\mathbf x \in \mathbf D$.  
Here we have assumed $f$ is defined over all of $\mathbf D$; for cost functions it often suffices to consider \eqref{eq:fcnRep} enforced over the feasible subspace.  

We next give a number of basic primitives from which Hamiltonians $H_f$ may be constructed, as well as more general operators needed for quantum optimization algorithms.

\subsection{Primitives and subroutines}\label{sec:primitives}

Here we 
develop an intermediate representation that is particularly useful when mapping classical discrete optimization problems into quantum algorithms. 
The representation is based on a small number of fundamental primitives, to which any classical function on or transformation of discrete variables may be mapped. If the user desires, some analysis 
of the problem and algorithm may then be performed at this intermediate level, before a particular qubit-based (or other) encoding is implemented in an automated way. 
This means one does not need to consider the particular encoding or hardware details until \textit{after} the ``quantization'' of the combinatorial problem, and constructions may in principle be easily transferred across different encodings and devices. We call this construction the Discrete Quantum Intermediate Representation (DQIR). DQIR then easily facilitates implementation of 
a wide variety of quantum algorithms such as the quantum approximate optimization algorithm and its generalization to the quantum alternating operator ans\"atze (QAOA)~\cite{farhi2014quantum,hadfield19_qaoa}, quantum annealing~\cite{albash2018adiabatic}, variational approaches~\cite{moll2018quantum}, and quantum imaginary time evolution (QITE)~\cite{mcardle2019ite,Motta19_qite}, as well as the novel algorithms of tomorrow. For the reader's benefit we briefly review some of these approaches in Sec.~\ref{sec:qalgs}.  

We begin by considering a single discrete variable. In our formalism, the $d_\a$ values of a classical \textit{discrete variable} $x_\a$ are mapped one-to-one to the $d_\a$ levels (labeled with integers $\{0,\cdots,d-1\})$ of a  \textit{quantum discrete variable} (or quantum variable) that can be abstractly conceptualized as a qu\textit{d}it~\cite{hadfield19_qaoa,sawaya20_dlev}. 
Throughout this paper, we label discrete variables with a Greek letter and their values with Latin letters. As an example, in a graph coloring problem, each node is mapped to its own quantum variable while the discrete color value corresponds to a level in the quantum variable. 
Though this work focuses on 
discrete variables with $d>2$, we emphasize that binary variables and problems 
are also 
subsumed by this framework.
%

\paragraph{Diagonal primitives.}
For a single discrete variable $x_\alpha$, the simple projector 
onto the discrete value that corresponds to level $k$ is
\begin{equation}
\mP\supr{k}_\a \equiv |k \ra\la k|_\a,
\end{equation}
which we call the \textit{indicator primitive} because it represents the function that is $1$ if and only if variable $x_\alpha$ is assigned $k$. 

For general single-variable functions we then define the 
\textit{value primitive}
%
\begin{equation}\label{eq:value_prim}
\mA_\a = \sum_{k=0}^{d-1} a_{k,\alpha} |k \ra\la k|_\a
\end{equation}
which 
diagonally applies 
an arbitrary scalar value $a_k$ to each 
level $k$. 

 We emphasize that although $\m A$ is constructed using $\m P$ and the set of all $\m P$ is contained in the set of all $\m A$, it is useful to think of both as primitives. This is because $\m P\supr{k}_\a$ is used as a marker to ensure that a 
 variable is in a particular state, whereas $\m A$ may be used as a drop-in replacement for a classical variable or function. 
 %
 
 Often $\m A_\alpha$ simply
returns the integer label $a_k = k$, an important case that we appropriately call the \textit{number operator} and denote
%
\begin{equation}
\m N_\a \equiv 
\sum_{k=0}^{d-1} k |k \ra\la k|_\a,
\end{equation}
which appears in the special case where the labels denote occupation number as in quantum physical systems \cite{sawaya20_dlev}. 
Other functions are similarly defined through the coefficients $a_{k,\alpha}$. 
%

The indicator and value primitives 
over different variables may be combined through linear combinations to represent any classical Boolean-valued, discrete-valued, or real-valued function. 
Properties of functions on binary variables are studied in \cite{hadfield18_repres}, many of which immediately generalize to the case of integer domains; see for instance~\cite[Ch. 8]{o2014analysis} for additional details.




%


\paragraph{Multi-variate functions and examples.}
DQIR builds all multivariate logic from single variable primitives.
We introduce Greek subscripts to label each quantum variable. Any multivariate operator may then be expressed as a sum of tensor products of local operators,
\begin{equation}\label{eq:sum_tens_prod}
\sum_j c_j \bigotimes_\a \m B_{j,\a}
\end{equation}
where $\m B_{j,\a}$ 
is a single-variable primitive, 
which includes both the diagonal case as well 
as operators built from the more general primitives 
we consider below. 
Note that 
the non-diagonal primitives and operators we consider 
do not in general satisfy $\m B_{j,\a}^\dagger=\m B_{j,\a}$. 

For the diagonal case, any 
function may be expressed as a weighted sum of Boolean-valued functions, 
which is a common 
form of problem cost functions, 
and so it is especially useful to be able to build up 
Hamiltonians representing complicated functions from simpler Boolean projectors. 
It is further useful to be able to compose them through standard logical operators in order to represent more complication Boolean formulas or circuits. 
The following expressions relating Boolean logic 
on binary functions and variables \cite{hadfield19_qaoa} to their resulting Hamiltonian representations 
directly extend to our more general discrete variable setting: 
\begin{align}\label{eq:bool_primitives}
 H_{\neg f} &= I - H_f         \\ 
 H_{f \Rightarrow g} &= I - H_f + H_f H_g  \\
 H_{f \land g} &= H_{fg} =  H_f H_g         \\ 
 H_{f \lor g} &= H_f + H_g - H_f H_g \nonumber  \\
 H_{f \oplus g} &= H_f + H_g - 2 H_f H_g     \\ 
 H_{af+bg} &= aH_f + bH_g \mspace{18mu} a,b \in \mathbb R   \nonumber 
\end{align}
where $f,g$ are arbitrary $\{0,1\}$-valued functions on $\mathbf D$ and $H_f\ket{\mathbf x}=f(\mathbf x)\ket{\mathbf x}$. A particular useful property we employ below is that $H_fH_g$ is identically zero when $f,g$ are $1$ on disjoint sets. Here functions acting on fewer than 
all $M$ variables are trivially extended to all of $\mathbf D$. 
The logical rules of \eqref{eq:bool_primitives} apply to the diagonal primitives above and easily generalize to higher-order multivariate expressions through composition. 
Cost functions are often expressed as sums of Boolean clauses, for example in constraint satisfaction problems,  which can then be directly mapped to a cost Hamiltonian via the linearity property of \eqref{eq:bool_primitives}.
One may similarly consider the case of complex coefficients $h=\sum_i a_i f_i$, $a_i\in \mathbb C$,  
though the corresponding 
operator may no longer be Hermitian; for example, one may decompose a unitary operator this way. 

Next we make use of \eqref{eq:bool_primitives} to write down Hamiltonians corresponding to some prototypical 
multivariate functions. 
A first important example 
is the equality operator
\begin{equation}\label{eq:EQ}
\eq(\alpha,\beta) = {\bigvee}_{a=0}^{d-1} \left ( \mP\supr{a}_\alpha \land \mP\supr{a}_\beta \right ) = \sum_{a=0}^{d-1} \mP\supr{a}_\alpha \mP\supr{a}_\beta 
\end{equation}
which vanishes when the values on variables $\alpha$ and $\beta$ are unequal, else returns $1$ (\textit{i.e.} acts as the identity) when they are the same. 
The case where variables $x_\alpha,x_\beta$ take values in different domains is easily handled by considering only their pairwise intersections in \eqref{eq:EQ}. The \textit{not equal} operator is then defined as $NEQ(\alpha,\beta)=I-\eq(\alpha,\beta)$. 
%
An example of a function with an arbitrary number of variables is the \textit{all equal} function over $M$ variables,
\begin{equation}
\tr{AEQ} = {\bigvee}_{a=0}^{d-1} \left ( \bigwedge_{\a}^{M} \mP\supr{a}_\alpha \right ) = \sum_{a=0}^{d-1} \prod_\a^{M} \mP\supr{a}_\alpha.
\end{equation}
%
The \textit{all different} function (see, \textit{e.g.} \cite{Booth21_constrprog}) on $M$ integer variables can be expressed as 
\begin{equation}
\tr{AD} = \prod_{\a \neq \b}^{M} (I - \eq(\a,\b)).
\end{equation}
A simple example of an integer-valued function is 
the \textit{count non-zero} function
\begin{equation}
\tr{CNZ} = M - \sum_{\a}^M \m P\supr{0}_\a.
\end{equation}
Another example, given a graph $G=(V,E)$ with variables as nodes, is the \textit{pairwise different} function
\begin{equation}
PD = \sum_{(\alpha \beta)\in E}NEQ(\alpha,\beta),
\end{equation}
which for a node coloring $\mathbf x$ counts the number of cut (differently colored) edges in $G$.

Clearly a wide variety of cost functions or constraints can be represented as quantum operators in this way; we explore some concrete application examples in Sec.~\ref{sec:problems}. 

\paragraph{Non-diagonal primitives.}
Naturally, in addition to classical functions, we also need to represent non-diagonal operators 
that facilitate shifting probability amplitude between different computational basis states. Thus we next introduce two additional classes of single-variable primitive.

We first define the one- and two-way 
\begin{equation}\label{eq:transfer_prim_oneway}
\m T^{(k \leftarrow l)}_\a \equiv |k \ra\la l|_\a,
\end{equation}
\begin{equation}\label{eq:transfer_prim_symm}
\tran{k}{l}{\a} \equiv |k \ra\la l|_\a + |l \ra\la k|_\a.
\end{equation}
%
These operators are useful for instance in designing mixers for QAOA. We deliberately do not restrict \eqref{eq:transfer_prim_oneway} to be Hermitian, in order to allow DQIR to represent general non-Hermitian operators that appear for instance in 
the analysis or implementation of algorithms such as QITE. However, in the most common use cases 
transfer primitives will appear as 
$\tran{k}{l}{\a}$.

The 
final single-variable primitive is the \textit{general local} operator 
%
\begin{equation}\label{eq:arbloc}
\m G_\a = \sum_{k=0,l=0}^{d-1} g_{kl} |k \ra\la l|_\a = \sum_{k,l=0}^{d-1} g_{kl} T^{(k \leftarrow l)}_\a.
\end{equation}
%
which 
generalizes the three previous primitives. Formula \eqref{eq:sum_tens_prod} may then be used to express 
any multivariate operator in terms of the single-variable primitives.

The four single-variable primitives are the most essential concepts of the DQIR workflow and result in a convenient unified quantum representation for any classical function on discrete variables and any transformations between classical states. 
Additionally, we emphasize that DQIR objects may be 
constructed and manipulated independently of qubit encoding choice or other lower-level concerns, and one may perform symbolic algebraic manipulations and analyses within DQIR. In the remainder of this subsection we demonstrate explicit constructions of several important classes of functions and operators. 

\paragraph{Single-variable 
reversible functions.}
Logically reversible functions are the essential building blocks of both classical reversible and quantum computing.  
For a binary-valued variable the only non-identity bijective (i.e., one-to-one and onto) function is $\{0,1\}\mapsto\{1,0\}$, which in qubit space can be implemented with the quantum gate $X$, and can be undone reapplying the same transformation. Here we generalize the quantum implementation of such single-variable reversible 
discrete functions for $\mathbb{Z}_d \mapsto \mathbb{Z}_d$ with $d \geq 2$. A bijective function on $d$ integers is just a permutation $\tau(k)$ of $d$ elements, $k \in \mathbb{Z}_d$. 
Any permutation $\tau$ on $d$ elements is representable as a unitary $U_{\tau}$. We write
\begin{equation}\label{eq:sgl-var-perm}
U_{\tau} = \exp(-i \frac{\pi}{2} H_{\tau})
\end{equation}
%
with Hermitian $H_{\tau}$, as any unitary matrix may be expressed as the exponential of a Hermitian. 
$H_{\tau}$ may be represented in DQIR with the help of transfer primitives.

Note that it may often be useful (for example when $H_{\tau}$ is not known or its exponential is resource-intensive) to consider a decomposition into simpler permutations such that $\tau(a) = \tau'(a)\tau''(a)\cdots$, where the exact implementation of each $\{\tau'(a),\tau''(a),\cdots\}$ is known. For example, considering that any permutation may be constructed from pairwise exchanges, one may implement $\tau(a)$ via individual transfer primitives:
\begin{equation}\label{eq:perm-by-pairs}
U_{\tau} = \tran{k}{l}{\a} \tran{k'}{l'}{\a} \cdots.
\end{equation}
where $\tran{k}{l}{\a}=\exp \left[ -i \frac{\pi}{2} \left( \tran{k}{l}{\a}  -I \right) \right]$ and I is the identity operator. 


\paragraph{Controlled instructions.} 
DQIR may be used to facilitate controlled quantum instructions as well, 
where a unitary $U=U(\phi)$ is applied conditioned on the variable $\alpha$ being set to value $k$,
\begin{equation}\label{eq:contrU}
   U'= \mP\supr{k}_\a \otimes U + (I-\mP\supr{k}_\a )\otimes I.
\end{equation}
This includes the case where $U=U(\phi)$ (and hence $U'=U'(\phi)$) are parameterized unitaries.
If 
the target unitary can be expressed as $U=e^{-i \phi H}$ 
then the operation
\begin{equation}\label{eq:ctrUexp}
    U'(\phi)=e^{-i \phi \mP\supr{k}_\a \otimes H}
\end{equation}
produces the desired controlled operation, conditional on variable $\a$ being in state $k$.

%

Using the observation that any Hamiltonian $H_\fint$ representing a $\{0,1\}$-valued Boolean function $\fint$ gives a projector~\cite{hadfield18_repres}, 
we generalize the control part of equation \eqref{eq:ctrUexp} from the indicator primitive to any multivariate Boolean-valued function $H_\fint=\fint(x)\ket{x}$ on discrete variables,
\begin{equation}\label{eq:SELECTmany}
U_\fint' = e^{-i \phi H_\fint \otimes H}
\end{equation}
where $H_\fint$ and $H$ act nontrivially on distinct sets of qubits, and 
$U_\fint' $ acts as $U$ when the function $\fint$ is satisfied by the control register variables, else as the identity.   

For multiqubit operators, the control function typically considered is the AND operation over a subset of variables (or their negations), as for example in multiqubit Toffoli gates~\cite{barenco1995}. The generalization to arbitrary functions on Boolean domains is considered in \cite{hadfield18_repres}. Our case of $d$-ary domains is much more rich, with a much larger set of possible single- and multi-variable Boolean expressions, for example, controlling on multiple states for each variable.   
Applying the rules of \eqref{eq:bool_primitives} it is relatively straightforward to construct controlled operators for a wide variety of commonly encountered conditional expressions. Moreover, unitaries corresponding to different controlled functions can be applied in sequence to generate multi-case controlled operators.    

\paragraph{Computing functions into registers.} 
Similarly, directly computing functions into registers is possible with 
operators constructed using DQIR representations, an important special case of multi-controlled instructions. 
For a Boolean-valued function $\fint(\mathbf x)$, we may compute its value in an additional qubit register as 
\begin{equation}
G_\fint \ket{\mathbf x}\ket{a} = |\mathbf x\ra|a \oplus \fint(\mathbf x)\ra 
\end{equation}
by applying the exponential
\begin{equation}
G_{\fint} = e^{-i \frac{\pi}{2} H_\fint \otimes (X-I) }
\end{equation}
where $X$ is the Pauli operator and $\oplus$ denotes addition modulo 2.


Here we show how to generalize this approach to computing powers of arbitrary bijective discrete functions $\tau:\mathbb{Z}_d \mapsto \mathbb{Z}_d$. We wish to implement 
$\tau(a)$ conditional on the result of an 
integer-valued function $f(\mathbf x)$ where $\mathbf x \in \mathbb Z_{d_1} \times \dots \times \mathbb Z_{d_p}$. 
The operator of interest is 
\begin{equation}\label{eq:perm_register}
e^{-i \pihalf H_\fint \otimes H_{\tau} }\ket{\mathbf x}\ket{a} = \ket{\mathbf x}|\tau^{\fint(\mathbf x)}(a)\ra,
\end{equation}
%
where $\tau^k(a)$ signifies $k$ repetitions of the bijective function $\tau(a)$. If $\fint(\mathbf x)$ is a binary-valued function, then its output determines whether to perform the permutation or not. If $\fint(\mathbf x)$ is an integer-valued function 
then the permutation may be applied multiple times. 
Notably, the set of operations \eqref{eq:perm_register} contains the subclass
\begin{equation}
W\ket{\mathbf x}\ket{a} = | \mathbf x \ra | a+p\fint(\mathbf x) \mod d \ra
\end{equation}
for arbitrary integer $p$, which we highlight because of its potential use for integer arithmetic. 

We emphasize that permutations are fundamental objects in reversible computation~\cite{shende2008_toffdecomp,saeedi2013synthesis} and hence \eqref{eq:perm_register} facilitates implementation of broad classes of functions. In particular arbitrary functions may be extended to reversible versions through the inclusion of ancillary variables~\cite{saeedi2013synthesis}.

\begin{table*}[]
\centering
\begin{tabular}{cccccc}
Base ten & SB & Gray & Unary & DW & BU$_{g=3}^\textrm{Gray}$ \\
0  & \texttt{0000}  & \texttt{0000} & \texttt{000000001} & \texttt{00000000} & \texttt{00 00 00 01}  \\
1  & \texttt{0001}  & \texttt{0001} & \texttt{000000010} & \texttt{00000001} & \texttt{00 00 00 11}  \\
2  & \texttt{0010}  & \texttt{0011} & \texttt{000000100} & \texttt{00000011} & \texttt{00 00 00 10}  \\
3  & \texttt{0011}  & \texttt{0010} & \texttt{000001000} & \texttt{00000111} & \texttt{00 00 01 00}  \\
4  & \texttt{0100}  & \texttt{0110} & \texttt{000010000} & \texttt{00001111} & \texttt{00 00 11 00}  \\
5  & \texttt{0101}  & \texttt{0111} & \texttt{000100000} & \texttt{00011111} & \texttt{00 00 10 00}  \\
6  & \texttt{0110}  & \texttt{0101} & \texttt{001000000} & \texttt{00111111} & \texttt{00 01 00 00}  \\
7  & \texttt{0111}  & \texttt{0100} & \texttt{010000000} & \texttt{01111111} & \texttt{00 11 00 00}  \\
8  & \texttt{1000}  & \texttt{1100} & \texttt{100000000} & \texttt{11111111} & \texttt{00 10 00 00}  \\
\end{tabular}
\caption{Integers 0 through 8 encoded into bits using standard binary, Gray, unary (one-hot), domain wall, and BU$_{g=3}^\textrm{Gray}$. The latter encoding refers to block-unary where each block uses the Gray code.}
\label{tbl:encs}
\end{table*}

\subsection{Actions within DQIR}

A primary advantage of DQIR is that 
it unifies the diverse landscapes of discrete problems and quantum algorithms under one 
operator representation, facilitating the design and deployment of automated tools that can be applied to suitably prepare 
arbitrary discrete problems for a quantum computer. Some such automated tools and actions are listed in Figure \ref{fig:prob-ir-enc}.

One purpose of an intermediate representation is to enable subroutines and analyses that are independent of the broader problem or algorithm type. At the simplest level, DQIR is useful as a way to cancel, collect, combine, and reorder terms in operators before one implements a qubit, or other, encoding. Because the terms are often fewer and simpler in DQIR than 
in subsequent lower-level 
representations, this will often reduce the complexity of the resulting algebra that needs to be performed 
(\textit{e.g.} compared to Pauli expression manipulation). Working at the DQIR level can also 
avoid redundant work when analyzing across multiple encodings.

Higher-level analyses may be performed in DQIR as well. 
For instance, once 
a Hamiltonian is constructed in DQIR one may
directly estimate its 
norm or other quantities of interest, 
as this may guide the choice of parameters such as time step size used in a quantum algorithm, or even guide the choice of algorithm itself. 
Analyzing the connectivity of the discrete variables, \textit{i.e.} considering the underlying graph properties of the problem, may be very useful at this level as well. 
For a given hardware choice, 
one may reorder the 
quantum variables within DQIR, for example using a clustering algorithm, as to ensure that minimal 
non-native interactions are 
needed in the quantum device. 
In some cases one may choose to reduce the size of the problem by solving only highly connected variables on the quantum computer \cite{gg21_divide,saleem21_divide}. 

\subsection{Lowering DQIR into hardware-relevant representations}\label{sec:dqir2qub}

While for simplicity we focus here on qubit-based digital hardware with all-to-all connectivity, it is straightforward to incorporate alternative or additional encoding layers into DQIR, including mappings that account for hardware topology limitations, as well as noise via the broad field of quantum 
fault tolerance. Similarly, one may consider qudit-based hardware, including the general case where  
the dit and variable dimension don't necessarily match.


\paragraph{Mapping to qudits.}
Here 
the general goal is to convert DQIR to a multi-qu\textit{d}it representation where 
variables are 
encoded with one or more qudits 
via an embedding
\begin{equation}\label{eq:to_qudits}
\mathbb Z_{d_1} \times \cdots \times \mathbb Z_{d_K} \, \xrightarrow{\m R^{enc}} \, \mathbb Z_{d_1'} \times \cdots \times \mathbb Z_{d_{K'}'}.
\end{equation}
In general the sizes of the variable domains may not be equal to the 
sizes of the encoded multi-qudit space; indeed the original domain is often strictly smaller than the embedded space. 
Such a dimension mismatch allows for one-to-many mappings, or requires that some states in the encoding domain be unused. We call an encoded computational basis state \textbf{valid} 
if it corresponds to a state in the original domain. 
\newcommand\mapsfrom{\mathrel{\reflectbox{\ensuremath{\mapsto}}}}
For example when a variable with $d=5$ values is mapped to qubits using the standard binary encoding, 
$|011\ra \;(\mapsfrom|k=3\ra)$ is a valid state while $|111\ra$ is an invalid state. Valid states may or may not be \textit{feasible} depending the particular problem at hand. For instance, consider the traveling salesperson problem encoded as a permutation problem (see equation \eqref{eq:x_permut}) on three cities using the standard binary encoding, such that the cities are labeled as $\{0,1,2\} \mapsto \{|00\ra,|01\ra,|10\ra\}$. We refer to a six-qubit state such as $|00\ra|10\ra|00\ra$ ($\mapsfrom |0\ra|2\ra|0\ra$) as valid but infeasible, because though it has a corresponding value in the original space (validity), it does not represent a permutation. Note that for unconstrained problems we may use the terms feasible or valid interchangeably. 


\begin{figure}
    \centering
    \includegraphics[width=.5\textwidth]{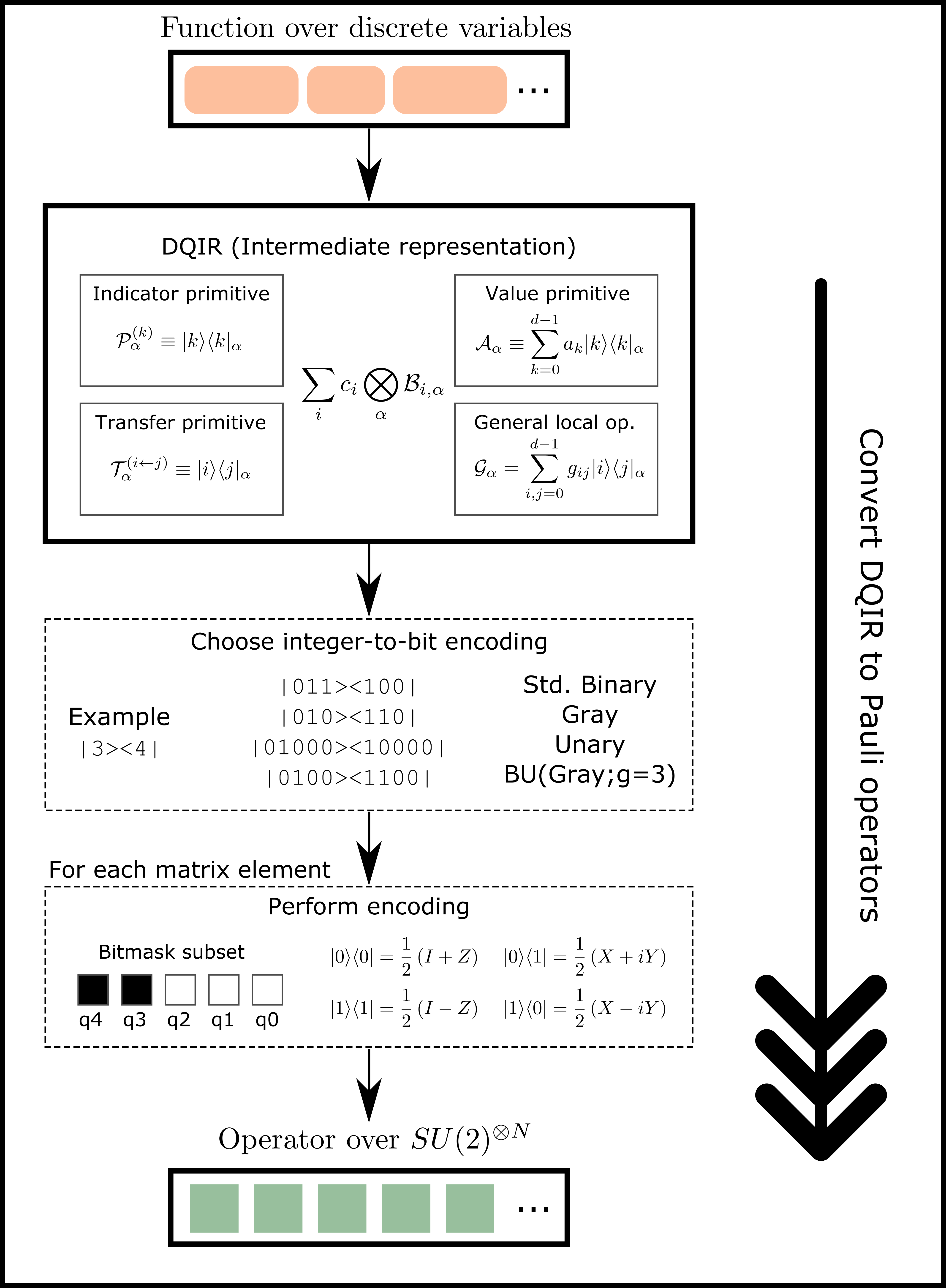}
    \caption{Flow diagram for converting from DQIR to qubit-based operators. This procedure is automated for any integer-to-bit mapping, of which four examples are shown. Values of the bitmask subset are shown in blue. A similar workflow would apply to compilation to other platforms such as qudit-based quantum computers.
    }
    \label{fig:ir-enc-detail}
\end{figure}

\paragraph{Mapping to qubits.}

In the remainder of this work 
we focus on the important special case of independently mapping each single discrete variable to 
qubits
%
\begin{equation} \label{eq:qubitEnc}
\mathbb Z_d \xrightarrow{\m R^{enc}} \mathbb Z_{2} \times \cdots \times \mathbb Z_{2},
\end{equation}
where for simplicity we 
will further assume that the mapping is injective (\textit{i.e.}, not one-to-many, which would be the case for example if each element were mapped to a larger subspace). We note however that a one-to-many mapping is possible as well and has been proposed in the context of QAOA for Max-$k$-Cut \cite{fuchs21_maxkcut}. We will consider a number of explicit examples of such qubit mappings, see Table~\ref{tbl:encs}. 
%
%
The encoding \eqref{eq:qubitEnc} of states also induces a mapping of DQIR primitives and expressions to qubit operators. 
Operators on qubits are commonly expressed as sums of products of Pauli operators and critically depend on the particular encoding scheme selected. Much of the following procedure has been given in a pedagogical way in previous work \cite{sawaya20_dlev} but here we give an overview, shown schematically in Figure \ref{fig:ir-enc-detail}. 
While here we focus on mapping to \textit{logical} (encoded) qubits, DQIR may be easily incorporated into quantum error correction schemes~\cite{lidar2013quantum} to derive operators at the \textit{physical} qubit (or qudit) level. 

We will consider several common qubit encodings drawn from the literature, summarized in Table~\ref{tbl:encs}. 
While some encodings require significantly more space (qubits) than others, on the other hand, within a given encoding and gate set a given operator may be much easier to 
implement than in another. 
For a given algorithm 
the choice of encoding presents immediate trade-offs between qubit count and circuit depth, as well as other measures or circuit complexity, though it is not clear generally how such resource trade-offs ultimately relate algorithm to performance, which is a 
rich but complicated topic beyond the scope of this work. 

Furthermore, it may sometimes be desirable to employ different encodings for different variables, a general strategy we call \textit{mixed encoding}. First, 
for heterogeneous problem domains 
with differing $d_\alpha$, 
different encodings may be optimal (in terms of circuit depth) for different variables, 
as will be demonstrated in Section \ref{sec:results}. Second, even when variables have the same value of $d_\alpha=d$, hardware constraints such as irregular connectivity or differences in individual qubit quality might lead to performance advantages from mixed encodings. 

Consider again the integer-to-bit encoding 
$\m R^{enc}$ 
of Eq. \eqref{eq:qubitEnc} which maps from a discrete value to an encoding-dependent number $N_q$ of bits, $R^{enc}: \mathbb Z_d \rightarrow \mathbb Z_2^{\otimes N_q}$. 
While there exist in principle exponentially many 
such encodings, common encodings may be broadly grouped in terms of their trade offs between space and depth overheads.   
In this work we consider the standard binary (SB), Gray, and unary (one-hot) encodings,
as well as a class of encodings that interpolates between them called \textit{block unary} \cite{sawaya20_dlev}, as shown in Table~\ref{tbl:encs}. 
These encodings have been previously studied 
in the context of resource advantages for physics and chemistry simulations~\cite{sawaya20_dlev}. 
For each variable the compact codes (SB and Gray)
require $\ceil{\log_2 d}$ qubits, unary requires $d$ qubits, and block unary interpolates between the two, requiring
$\ceil{\frac{d}{g}}\ceil{\log_2(g+1)}$ qubits where $g$ is an integer parameter (assuming a compact code is used for the local encoding of each block). 
An alternative unary approach called the \textit{domain wall} (DW) encoding \cite{chancellor19} uses one fewer qubit than one-hot; DW has been shown to yield significantly superior algorithm performance than one-hot in multiple contexts \cite{berwald2023understanding,plewa2021variational}. 


\paragraph{Bitmask subsets.}


%
\begin{table*}[]
\centering
\begin{tabular}{ccccc}
Primitive & Compact (SB,Gray,\dots) & Unary & Domain Wall & BU$_{g=3}$ \\
$\m P\supr{0}_\a \equiv \ket{0}\bra{0}$  & \texttt{****}   & \texttt{\_\_\_\_\_*} & \texttt{\_\_\_\_*} & \texttt{\_\_ **}  \\
$\m P\supr{1}_\a \equiv \ket{1}\bra{1}$  & \texttt{****}   & \texttt{\_\_\_\_*\_} & \texttt{\_\_\_**} & \texttt{\_\_ **}  \\
$\m P\supr{2}_\a \equiv \ket{2}\bra{2}$  & \texttt{****}   & \texttt{\_\_\_*\_\_} & \texttt{\_\_**\_} & \texttt{\_\_ **}  \\
$\m P\supr{5}_\a \equiv \ket{5}\bra{5}$  & \texttt{****}   & \texttt{*\_\_\_\_\_} & \texttt{*\_\_\_\_} & \texttt{** \_\_}  \\
$\tranasymm{1}{2}{\a} \equiv \ket{1}\bra{2}$  & \texttt{****}   & \texttt{\_\_\_**\_} & \texttt{\_\_***} & \texttt{\_\_ **}  \\
$\tranasymm{2}{5}{\a} \equiv \ket{2}\bra{5}$  & \texttt{****}   & \texttt{*\_\_*\_\_} & \texttt{*****\_} & \texttt{** **}  \\
\end{tabular}
\caption{Bitmask subsets for various 
single-variable primitives on a discrete variable $\a$ with $d=6$ states $\ket{0}$ through $\ket{5}$. `\texttt{*}' and `\texttt{-}' denote qubits that are inside and outside the bitmask subset, respectively, where the qubit ordering is $\{\cdots, 2, 1, 0\}$. Note that the first four rows correspond to diagonal elements while the last two rows correspond to off-diagonal elements. For each 
primitive DQIR-to-qubit mapping procedure need consider only qubits in the bitmask subset. Multi-variate operators are built from single-variable primitives as in equation \eqref{eq:sum_tens_prod}.  
}
\label{tbl:bitmask}
\end{table*}
%

%
Consider a DQIR single-variable primitive $\tranasymm{k}{l}{} \equiv \ket{k}\bra{l}$. 
As 
described in previous work \cite{sawaya20_dlev}, in order to take advantage of the sparsity of non-compact encodings one 
can introduce the concept of a bitmask
subset $ \m C^{\tr{enc}}(\tranasymm{k}{l}{}) \equiv \m C^{\tr{enc}}(\{k,l\})$,
the subset of bit (qubit) indices 
for which the resulting encoded qubit operator acts nontrivially. 
The bitmask subset is a useful concept because it 
facilitates automated implementation of encodings beyond just standard binary and Gray, in a way that does not operate on more qubits than are strictly required. Qubits not in the bitmask subset may safely be ignored. Hence the size of $\m C^{\tr{enc}}(\{k,l\})$ determines the 
qubit locality (number of qubits on which it operates nontrivially) of the encoded primitive. Examples of bitmask subsets for various elements and encodings are shown in Table \ref{tbl:bitmask}. 

When considering diagonal elements $\m P\supr{k} \equiv \ket{k}\bra{k}$, $\m C^{\rm{Compact}}(\;\{k\}\;)$ for compact codes consists of all bits in the quantum variable, while $\m C^{\tr{Unary}}(\{k\})$ for integer $k$ is simply the singleton set 
$\{k\}$. Asymptotically, the size of the bitmask subset for block unary is in-between the sizes of those for compact and unary, \textit{i.e.} 
$$
|\m C^{\tr{Unary}}(\;\{k\}\;)| < |\m C^{\tr{BU}}(\;\{k\}\;)| 
$$
$$
\leq |\m C^{\tr{Gray}}(\;\{k\}\;)| = |\m C^{\tr{SB}}(\;\{k\}\;)|,
$$ 
though for smaller $d$ this trend does not always hold.
%
%
Performing the integer-to-bit encoding for each 
primitive in the computational basis yields 
\begin{equation}
\tranasymm{k}{l}{} \equiv | k \ra\la l | \rightarrow \bigotimes_{i \in \m C(\{k,l\})} | b_i \ra\la b'_i |_i
\end{equation}
where $b_i$ and $b_i'$ are the bit values for qubit $i$ resulting from the mapping, with implicit identity factors on the qubits outside of the bitmask $\m C(\{k,l\})$. The following identities may then be used to convert the 
right-hand side primitives to the Pauli qubit operators $X,Y,Z$: 
\begin{equation}\label{eq:ij2pauli}
\begin{split}
|0\ra\la 1| &= \tfrac12 ( \sx + i\sy ); \; |1\ra\la 0| = \tfrac12 ( \sx - i\sy ); \; \\
|0\kb 0| &= \tfrac12 (I + \sz); \; |1\kb 1| = \tfrac12 (I - \sz).
\end{split}
\end{equation}
%

We note that in the domain wall (DW) encoding, a particular integer often corresponds to many qubits being in the 1 state (see Table \ref{tbl:encs}). Because of this, the bitmask subset for an off-diagonal element $\tranasymm{k}{l}{}$ for $l>k$ is instead $\m C^{\tr{DW}}(\{k,l\}) = \{(k-1),k,k+1,\cdots,l-1,(l)\} $, where $l$ is included only if $l \neq d$ and $k-1$ is included only if $k\neq0$. 
Thus DW yields a larger bitmask subset than in the one-hot case and often larger than in the compact codes. However, as long as all transition primitives operate only on nearest-integers---which is the case with discrete mixers typically considered for QAOA---DW will lead to at most 3-local operators regardless of $d$. 
Notably, DW has been shown to provide advantages over one-hot in some cases, including fewer one-qubit operators in the Pauli basis 
\cite{chancellor19}. 
Though we do not include 
resource analysis for the DW encoding in our numerics of Sec.~\ref{sec:results}, we speculate that circuit depths will be roughly similar to the unary (one-hot) case for the specific (and somewhat narrow) subroutine we analyze, i.e. for operator exponentiation. 
It is important to point out that one must analyze a full algorithm end-to-end in order to determine which encoding performs best for a given application. Notably, for some applications DW has been shown to out-perform one-hot in quantum annealing ~\cite{berwald2023understanding} and QAOA \cite{chen2021performance}.

\paragraph{General remarks.}
As mentioned, 
there is often significant 
trade-offs between encodings in the required space (number of qubits) and  number of operations. 
Although a unary approach requires more qubits, it often leads to a shorter circuit depth. Consider a variable with $d=16$. While unary codes require 16 qubits, compact codes require 4 qubits---but compact encodings usually yield an operator with more terms that additionally have a higher average Pauli weight. However, 
on the other hand there exist domains and problems for which a compact encoding is most efficient both in terms of space and operations counts, as discussed below.

Some conceptual results relevant for matching an application with an encoding have been 
studied previously \cite{sawaya20_dlev,sawaya20_connect}, where the number of entangling gates (not the circuit depth) was determined for various physics and chemistry applications. Here we summarize some of the previous findings. First, a lower Hamming distance between two bit strings leads to a Pauli operator with fewer terms. One direct implication of this is that the Gray code is often more efficient than SB, especially when implementing tridiagonal operators. Second, SB is often the optimal choice (outperforming even unary) for diagonal operators that we call diagonal binary-decomposable (DBD) operators, defined as operators than can be encoded in standard binary as $a_0 Z_0 + a_1 Z_1 + \cdots$ where $a_i$ is a real scalar.
Third, though it may seem that BU would yield operations counts in-between unary and compact, in physics applications it is often (though not always) inferior to both. This is because the bitmask subset is unfavorably large when two integers are present on different blocks. 



After  qubit operators have been obtained, the final compilation steps involve mapping the problem to a particular hardware. 
This implementation will be based on the native gate set, hardware topology, and possibly an error mitigation or correction procedure (Figure \ref{fig:big-flow}). If the goal is to determine a desired encoding for a given operator and set of hardware, then one may run through the compilation pipeline for many encodings before comparing resource counts such as gate counts, circuit depth, qubit counts, or approximate error bounds. In this manner, one may determine the most resource-efficient encoding for a given quantum device.

Notably, there may be circumstances under which one would convert between encodings in the middle of the quantum algorithm. This has shown to decrease required quantum resources in quantum simulation of some physics and chemistry Hamiltonians \cite{sawaya20_dlev} and is worth exploring toward novel approaches for combinatorial problems as future work.



\section{Algorithm components in DQIR}\label{sec:algo_compon}



\begin{figure*}[t]
    \centering
    \includegraphics[width=.98\textwidth]{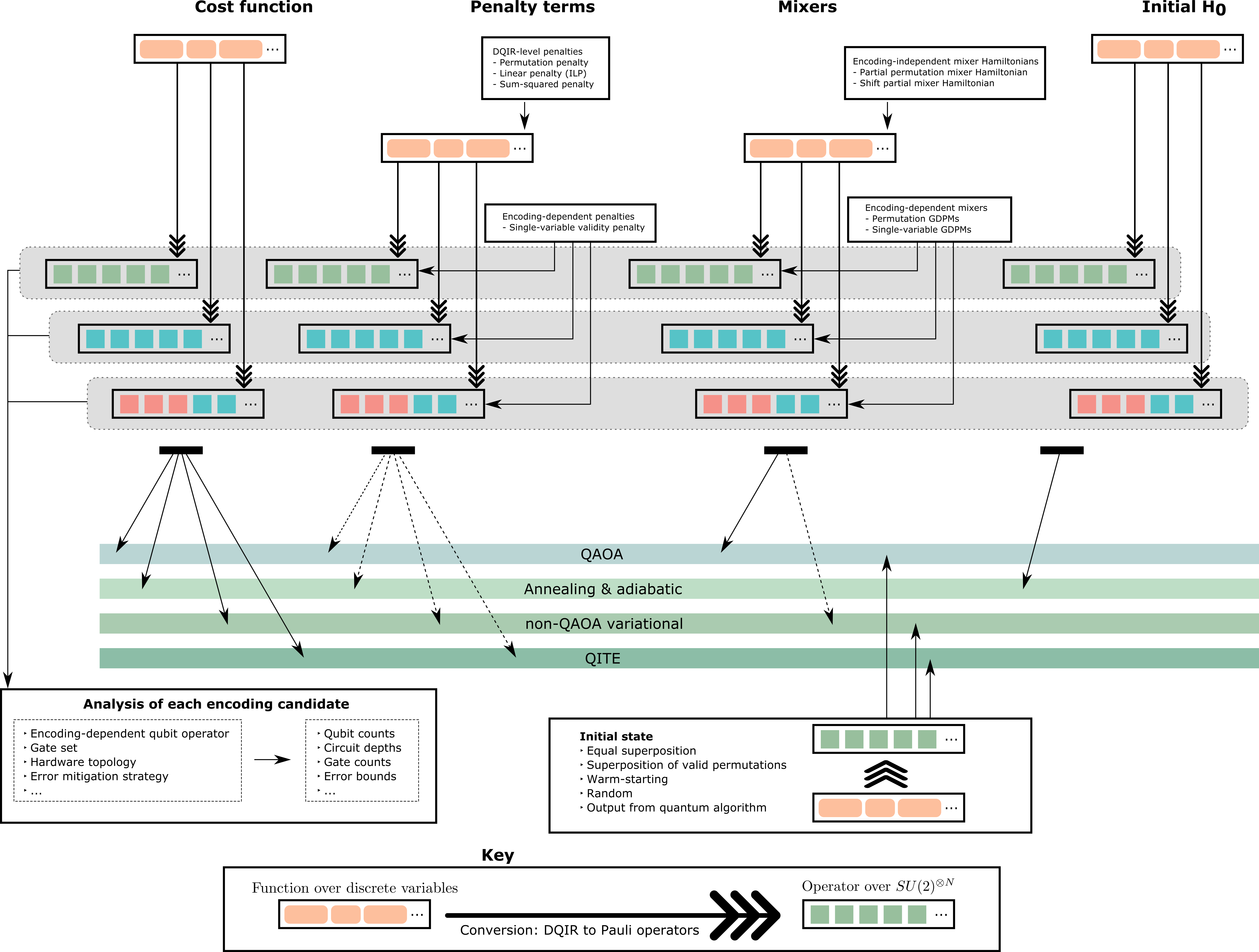}
    \caption{Flow diagram showing problem preparation for solving discrete combinatorial problems on quantum computers. Different colored squares denote different qubit encodings. Solid blue lines denote components that are required for a given algorithm class; dotted blue lines denote components that may sometimes be used for a given algorithm.}
    \label{fig:big-flow}
\end{figure*}


Here we provide a broad overview of various algorithmic components that are commonly required for 
tackling discrete optimization problems with existing quantum 
approaches, 
from the perspective of DQIR. 
We introduce several novel subroutines while striving to identify 
best practices and scenarios under which some algorithmic approaches are more 
advantageous than others. The flow chart in Figure \ref{fig:big-flow} guides the discussion. We again emphasize that as the ultimate power of quantum computers for combinatorial optimization remains a deep 
and active open research area, we avoid making claims regarding algorithm performance as much as possible and instead focus on tangible metrics such as comparisons of required resources for specific approaches and subroutines.

\subsection{
Quantum approaches to combinatorial optimization}\label{sec:qalgs}

We first 
motivate our results by briefly summarizing four prototypical classes of algorithms applicable to 
estimating 
low-energy eigenvalues and eigenvectors of a problem cost 
Hamiltonian $H_C$ (i.e., obtaining approximate classical solutions).  These are quantum annealing and adiabatic quantum optimization (AQO), QAOA (the quantum approximate optimization algorithm or, more generally, the quantum alternating operator ansatz), non-QAOA variational approaches, and finally approaches related to quantum imaginary time evolution (QITE). In each approach care must be taken to deal with any hard constraints, and we address several methods for doing so in detail. While a number of 
other approaches exist in the literature, and new approaches are frequently proposed in this 
rapidly developing field, the primitives required are typically similar to the 
ones we consider, so we uses these algorithms to illustrate the utility of our 
results for both current and future quantum methods. 
In particular, DQIR facilitates seamless extension 
in each approach to discrete variables of arbitrary $d_\alpha\geq 2$. 

Different quantum algorithms require different components and inputs to be produced from DQIR, as shown in Figure~\ref{fig:big-flow}. Although each of the mentioned algorithms may be 
employed as exact solvers, we consider them more generally in 
the context of approximate optimization, as quantum computers are not believed able to efficiently solve NP-hard problems. 
Moreover, in most cases rigorous performance bounds appear quite difficult to obtain so these algorithms can be considered as heuristic approaches, especially in the setting of near-term quantum hardware; see e.g. \cite{hadfield19_qaoa} for additional discussion. 

Again assume we are given a cost function on a discrete variable domain to minimize, suitably encoded as a cost Hamiltonian $H_C$, and possibly subject to a set of hard constraints, as defined in Section~\ref{sec:dqirProblemDefn}. In each of the algorithms considered below we seek to prepare a quantum state 
with at least non-negligible support on low cost states such that repeated state preparation and computational basis measurement yields such a solution with high, or at least non-negligible, probability. Note that this definition subsumes both the special cases of exact optimization (requiring the true optimum solution), as well as exact algorithms (that succeed with probability very close to $1$).





\paragraph{Annealing \& Adiabatic Quantum Optimization.}
In AQO 
as well as (closely related) quantum annealing protocols~\cite{johnson2011quantum,albash2018adiabatic,hauke2020perspectives} one begins in the ground state of a ``driver'' Hamiltonian $H_0$ for which said state is easy to prepare on a quantum computer, before gradually 
turning on the cost Hamiltonian $H_C$:
\begin{equation} \label{eq:aqo}
H(t) = (1-s(t))H_0 + s(t)H_C,
\end{equation}
where $s(t)$ is 
a suitable annealing schedule starting at 0, varying continuously, and terminating at 1 for some sufficiently large $t=T\gg 1$. The two primary choices in the algorithm design are the annealing schedule and~$H_0$. There have been a number of exciting recent innovations 
to the quantum annealing protocol
in terms of both hardware and theory, in particular more advanced annealing schedules accommodating so-called \textit{pause}~\cite{izquierdo2021ferromagnetically} and \textit{reverse}~\cite{venturelli2019reverse} features, among others, as well as novel 
hardware topology~\cite{dattani2019pegasus} and encodings~\cite{lechner2015quantum,chen2021performance}. 
Also notable are non-adiabatic annealing methods that may for example make use of environmental noise \cite{sarandy2005adiabatic,
amin2008thermally,
boixo2013experimental,
kechedzhi2016open,
passarelli2022standard}. 
Though this is a natural procedure for analog quantum devices (\textit{i.e.} quantum annealers), one may use Hamiltonian simulation algorithms 
such as 
Trotterized product formulas 
to approximately perform AQO 
on gate-based quantum computers, in terms of alternating \lq\lq bang-bang\rq\rq\ evolutions under $H_0$ and $H_C$.  
If we further relax the requirement that the sequence of discretized evolution 
must closely match the 
adiabatic one we naturally arrive at the QAOA family of parameterized quantum circuits, as we discuss next. We note that compilation of problems to actual quantum annealing hardware is a rich topic with quite distinct concerns from the gate model setting~\cite{zbinden2020embedding,konz2021embedding}.

\paragraph{Quantum alternating operator ansatz.}
In QAOA~\cite{hogg2000quantum,farhi2014quantum,hadfield19_qaoa}
one constructs a quantum circuit that alternates between 
applications of the so-called phase and mixing operators, applied to a suitable, efficiently preparable initial state:

\begin{equation} \label{eq:QAOA}
U_{mix}(\beta_p) e^{-i \gamma_p H_C} \cdots U_{mix}(\beta_1) e^{-i \gamma_1 H_C} |\psi_0\ra
\end{equation}
%
Each layer $j$ uses 
parameters $\gamma_j$ and $\beta_j$, $j=1,\dots,p$.  
The phase operator $e^{-i\gamma H_C}$ corresponds to time evolution under $H_C$ for a time $\gamma$. 
The mixing subroutine may be similarly implemented as the exponential of a mixing Hamiltonian $H_{mix}$, or more generally as 
some suitable parameterized unitary operator $U_{mix}(\beta)$ that meets certain design criteria discussed below. 
The algorithm parameters may be predetermined by analytic, empirical, or other means~\cite{wang2018quantum,bapat2018bang,zhou2020quantum,shaydulin2021classical,akshay2021parameter,streif2020training,verdon2019learning,wilson2021optimizing,magann2021feedback,brady2021optimal,wurtz2022counterdiabaticity}, or determined variationally using a hybrid quantum-classical search procedure~\cite{cerezo2021variational}. 
QAOA is 
inspired by but distinct from 
adiabatic algorithms, in that 
while in certain limits the QAOA state \eqref{eq:QAOA} can closely approximate the adiabatic evolution of \eqref{eq:aqo} 
\cite{farhi2014quantum}, for different parameters the resulting evolution can be significantly different, especially at a small or moderate number of layers~$p$.
Moreover, in general QAOA is not restricted to start in the ground state of the mixer. 
Even more so than the case of quantum annealing, a number of variants to QAOA have recently been proposed, see for instance~\cite{bartschi20_permut,bravyi2020obstacles,egger20_warm,wurtz2021classically,Liu21_layer_vqe}.

\paragraph{Variational 
quantum circuits beyond QAOA.}
Here we consider, broadly, more general classes of parameterized quantum circuits than those of QAOA. 
We use the term ``non-QAOA ansatz'' to refer to any 
such circuits that don't
strictly fit the definition of QAOA given above. 
Parameters may again be determined in general through variational optimization, or set through analysis or other means in specific cases. Here the quantum state ansatz may not depend on the cost function, or may but in a different way than in \eqref{eq:QAOA}.  
Relaxing the ansatz design  
gives greater flexibility and may help in fitting a quantum algorithm into the limited achievable circuit depths of early generation hardware, which may include hardware-tailored ansatz~\cite{moll2018quantum}. In principle one may use a short depth circuit with many more parameters than QAOA, at the expense of a much more challenging parameter setting task~\cite{mcclean2018barren,cerezo2021variational}, or include more complex circuit features \cite{zhu20_adapt_qaoa,Liu21_layer_vqe}. 
Similarly to the QAOA case, where one has freedom in the design of the mixing operator, 
one may typically trade-off between quantum resources, and classical pre- and post-processing requirements in 
designing an effective variational ansatz for a given class of problems.   


\paragraph{Imaginary time evolution.}
Quantum imaginary time evolution (QITE) algorithms~\cite{Motta19_qite,mcardle2019ite} determine and implement an approximation of the \textit{real} operator $\exp(-\beta H_C)$ to create the state
\begin{equation} \label{eq:qite}
|\psi(\beta)\ra 
\,\propto \,\exp(-\beta H_C) |\psi_0\ra\, = (\exp(-\tfrac{\beta}{k} H_C))^k  |\psi_0\ra,
\end{equation}
up to normalization, 
on a quantum computer. If such a state could be prepared for sufficiently large $\beta$ and with sufficient fidelity then we would be guaranteed to find a ground state of $H_C$ (assuming the initial state $\ket{\psi_0}$ has support on such states). However, as the QITE  operator is 
far from unitary for non-negligible $\beta$, 
it cannot be simultaneously implemented efficiently, exactly, and deterministically in general on quantum hardware. (Otherwise, for instance, quantum computers could efficiently solve NP-hard problems which is widely believe to not be the case; this is easy seen considering the initial state corresponding to a uniform superposition of bit (dit) strings.) Hence, after decomposing into Trotter steps each of small duration $\beta/k\ll\beta$ as indicated in \eqref{eq:qite}, QITE algorithms iteratively employ a hybrid quantum-classical procedure to determine a suitable local unitary approximation for each subsequent step. 
The procedure is expensive partly because as originally proposed \cite{Motta19_qite} each time step requires many iterations of quantum state tomography on a subset of the qubits that grows with each step. 
Understanding both the performance and limitations of QITE and related approaches remains an open and active research direction, in particular for the specific setting of combinatorial optimization, and especially what is achievable with near-term devices or polynomially-scaling resources more generally.  

We next turn to 
methods for adapting these approaches to problems with hard constraints, which we extend to our discrete variable setting.  Two primary strategies in the literature for 
dealing with hard constraints are penalty terms and constraint-preserving mixers, which we consider in turn. We further propose a hybrid approach that combines these two methods in Section~\ref{sec:pen_exch}.

\subsection{Penalties}\label{sec:penal}



Here we consider penalties, which are additional terms (diagonal operators) directly added to an existing cost Hamiltonian to produce an effective cost function that penalizes with added cost any violations of the hard constraints,
\begin{equation}
H_{\rm{eff}} = H_C + \chi_1 F_1 + \chi_2 F_2 \cdots,
\end{equation}
such that the low-energy states of $H_{\rm{eff}}$ correspond to the low-energy \textit{feasible} states of $H_C$. 
%
Each penalty term $F_j$ 
represents a suitable (usually non-negative) classical function, 
and
comes with a sufficiently large 
penalty weight~$\chi_j>0$
note that choosing optimal 
penalty weights is nontrivial and depends on the context and particular problem~\cite{izquierdo2021ferromagnetically}.  
Because finite-weight penalties do not strictly preserve the feasible subspace (\textit{i.e.}, transitions between feasible and infeasible or invalid states are possible, as opposed to the mixer-based approach of Sec. \ref{sec:mixers}), it is 
generally necessary to introduce a simple post-processing step of discarding invalid or infeasible
bit strings returned, or else attempting to `correct' them with a suitable classical procedure---for example a simple approach would be to correct an infeasible or invalid output state by finding the closest feasible string. 
In general there may be different possible ways to construct suitable penalty terms, and different choices come with different resource tradeoffs.


In terms of the algorithms of Section~\ref{sec:qalgs}, penalty terms are the standard approach in AQO for dealing with hard constraints. For QAOA, penalty terms may in principle be employed similarly, however, they may be much less effective \cite{hadfield19_qaoa,wang20_xymixers} because, as mentioned, in various parameter regimes QAOA may not resemble an adiabatic evolution such that penalty terms may not have the desired effect on the quantum state evolution. This observation 
hints at the alternative constraint-preserving mixer approach we consider in Section~\ref{sec:mixers}. 
Penalty terms may be similarly utilized in more general variational algorithms beyond QAOA, where similar consideration apply. 
Finally, for QITE, penalties (or another suitable approach) are necessary as without them 
the algorithm may converge to a wrong (i.e., infeasible or even invalid) state. 
For each approach, we 
note the distinction between including penalty terms within the quantum circuit or protocol directly, versus including it indirectly via the objective function to be optimized (typically the expectation of 
a cost Hamiltonian) in determining the algorithm parameters; the former can be seen as modifying the algorithm, where the latter is effectively a post-processing step. 

We distinguish the two 
most important types of penalty terms into two categories: discrete-space (DQIR-level) penalties and encoding-dependent (qubit-level) penalties. 

\subsubsection{DQIR-level penalties}
DQIR-level penalties are used to penalize violations of any of the unencoded classical problem's constraints of equations~\eqref{eq:equal_constr} and~\eqref{eq:inequal_constr}, \textit{i.e.} they are used to enforce 
\textit{feasibility} as specified by the problem input of the algorithm dynamics and output. Here we provide several penalty constructions 
for a number of commonly occurring examples of hard constraints. 
For simplicity of presentation we will assume uniform variable domains $D_\alpha=D_\beta$ 
of equal cardinality $d:=d_\alpha=d_\beta$ for each variable throughout; in most 
cases the generalization to arbitrary variable domains $D_\alpha\neq D_\beta$ is 
straightforward. Similarly, each primitive 
is easily restricted, as desired, to the case of acting on only a particular subset of the problem variables.

First, recalling equations~\eqref{eq:x_permut} and~\eqref{eq:perm_space} we define the pair permutation penalty as
\begin{equation}
F_{\rm{perm}} = 
\sum_{\a \neq \b} \sum_{k=0}^{d-1} \mP\supr{k}_\a \mP\supr{k}_\b,
\end{equation}
which is non-zero 
on states for which a discrete value 
occurs more than once, 
and so penalizes integer strings that don't encode 
permutations. Here we have employed the indicator primitives~$\m P$.

Next, a commonly encountered linear constraint is that all variables in a given set must sum to some constant $D$ (\textit{i.e.} when some such quantity must be preserved), for which one may use the squared-sum penalty 
\begin{equation}\label{eq:f_sum}
\begin{split}
F_{\text{sum}}(\textbf{x}) &= 
\left( \sum_\alpha \sum_k a_k \m P\supr{k}  - D \right)^2 \\
&= 
\left( \sum_\alpha \mA_\a - D \right)^2.
\end{split}
\end{equation}
Here squaring is used to ensure that any states violating the constraint are assigned higher energy by 
the penalty than those that do and is a common technique in penalty term design. 
We employ $F_{sum}$ in Sec.~\ref{sec:problems} below 
for the portfolio rebalancing problem.

General linear constraints are a further important constraint class 
that come in the form of 
inequalities such as $3x_\a + 2x_\b \leq b$, where 
the left-hand side is a weighted sum of a discrete variables and $b$ is a constant. 
In general these constraints yield a rectangular matrix $A$ such that $\sum_{i\a}A_{i\a}x_\a \leq b_i$. These linear constraints lead us to define penalty operators 
\begin{equation}\label{eq:f_ilp_i}
F_{\text{lin}}^{(i)} = 
\sum_{ A_{i\a} x_\a+ A_{i\b}x_\b+\cdots > b_i}^{x_\a < d_\a, x_\b < d_\b, \cdots}  
\m P\supr{x_\a}_\a \m P\supr{x_\b}_\b \cdots ,
\end{equation}
where the number of indicator primitives $\m P$ in the product is equal to the sparsity of row $i$. We further define 
\begin{equation}\label{eq:f_ilp}
F_{\text{lin}} = \sum_i F_{\text{lin}}^{(i)}
\end{equation}
where 
the $b_i$ are constants. We stress that a variable is included in the product of equation \eqref{eq:f_ilp_i} only if $A_{i\a}$ is non-zero.
The number of terms in the final encoded operator for penalty \eqref{eq:f_ilp_i} is heavily dependent on the sparsity of row $i$ of $A$;  
for many encodings the number of terms in the qubit-encoded operator scales exponentially with the number of nonzero elements in row $i$. Hence if even one row of $A$ is not sparse, using $F_{lin}$ will usually not be an efficient approach. In such cases, the introduction of a slack variable may be a preferable route \cite{hadfield19_qaoa,bako2022funcqaoa}. See Section \ref{sec:problems} for more discussion of $F_{\text{lin}}$ in the context of integer linear programming (ILP) problems. 

A wide variety of other useful constraints and penalties are possible and may be implemented at the DQIR level, in particular by directly applying the techniques of Section~\ref{sec:primitives}; we do not attempt an exhaustive enumeration here. We next turn to constraints and penalties that arise only after a lower-level encoding choice has been made.    

\subsubsection{Encoding-dependent penalties}

Unlike DQIR-level penalties, 
encoding-dependent constraints and penalties are those used when 
the target encoded space of the mapping \eqref{eq:to_qudits} also includes invalid variable assignments, \textit{i.e.} some encoded states that do not correspond to a state in the original domain~$\mathbf D$. 
Given such a fixed encoding, in some cases we can define suitable penalty terms at the DQIR level, then further compile these terms by applying the encoding to them; in other cases contextual lower-level knowledge can be utilized to derive suitable penalty terms. 

As this work 
primarily considers encodings for which each variable is mapped to its own set of qubits, here we define only \textit{single-variable 
validity penalties} which for a given variable $x_\alpha \in D_\alpha$ takes the form
\begin{equation}\label{eq:ss}
F_{SS} = 
\sum_{k \not\in \m S} \mP\supr{k}
\end{equation}
where $\m S:=\mathcal R_{enc}(D_\alpha)$ 
are the encoded assignments of $x_\alpha$ 
such that the sum is taken over any invalid states \cite{glos20_space_eff}. 
This penalty is intended for use primarily with compact codes, because for non-compact codes (e.g. unary) equation \eqref{eq:ss} requires a very large number of terms. For example, if one is using SB to encode a $d=3$ variable into two qubits, then one may 
impose a penalty cost on the 
qubit state $|11\ra$, which is not a valid configuration as it is not contained in $\{\ket{00}, \ket{01}, \ket{10}\}\;( \mapsfrom \{0,1,2\})$. 


The approach presented here may be extended
to more general multi-variable encodings and error correcting codes, which is often relatively straightforward on a case-by-case basis. Hence working at the DQIR level provides greater flexibility if the underlying hardware or encoding is later changed.

\subsubsection{Leakage} \label{sec:leakage}
%
Here we propose a simple measure that quantifies deviation from the desired feasible subspace, and as applicable both DQIR-level and encoded quantum states.
Given such 
a 
quantum state $\ket{\psi}$, we define its feasible outcome component as the probability of a measurement in the computational basis returning a feasible solution
\begin{equation}
    \mathcal{F}(\ket{\psi})=\la\psi|  \hat P_{\rm{feasible}} |\psi\ra,
\end{equation}
where $\hat P_{\rm{feasible}} := \sum_{\mathbf x \in \mathbf x_{\rm{feasible}}} | \mathbf x \ra\la \mathbf x |$ 

Hence, at the end of a quantum algorithm 
(for example, one that employs penalty terms such as AQO) the final quantum state produces a feasible outcome 
with probability $\mathcal{F}(\ket{\psi_f})=\mathcal{F}(
U\ket{\psi_0})$. 

Similarly, given a 
unitary operator~$U$ 
and a feasible state $\ket{\psi_0}$ 
(i.e., $\hat P_{\rm{feasible}}\ket{\psi_0}=\ket{\psi_0} )$,
we define \textit{leakage} due to $U$ with respect to $\ket{\psi_0}$ as the probability 
of a transition to a state that is either infeasible (valid but violates some constraints $g_k$) or even invalid 
(does not correspond to a state in the domain $\mathbf D$),  
\begin{equation}\label{eq:leakage}
\begin{split}
\mathcal{L}(U,\ket{\psi_0}) = 1-\mathcal{F}(U\ket{\psi_0}) \\
= 1 - \la\psi_0|
U^\dag(\beta) \hat P_{\rm{feasible}} 
U(\beta)|\psi_0\ra.
\end{split}
\end{equation}

Leakage is a critical consideration when one desires to construct operators that shift probability amplitude between feasible states only, such as driver operators in AQO and similarly, mixers in QAOA.     
We note here two 
contrasting examples. 
First, problems on an unconstrained domain $\mathbf D$ without any input hard constraints $g_k$ correspond to $\hat P_{\rm{feasible}} = \prod_\a^M (\sum_{k=0}^{d-1} \mP\supr{k}_\a)$, i.e., the only possible leakage is to invalid 
states that may arise from encoding choice; if no such invalid states exist then $\hat P_{\rm{feasible}}$ gives the identity operator.   
Second, under our conventions permutations problem correspond to $\hat P_{\rm{feasible}} = \sum_{\pi_k \in S_M} | \mathbf x_{\pi_k} \ra\la \mathbf x_{\pi_k} |$, where in general leakage to both infeasible or invalid states may occur.

Finally, we emphasize that we may apply \eqref{eq:leakage} in either case of evaluating individual operators, or an overall quantum algorithm. For the latter case, increased leakage typically relates to increased classical resources in terms of additional circuit repetitions required to compensate for 
diminished success probability. Moreover, while for simplicity we do not distinguish here between leakage to invalid versus valid but infeasible subspaces, 
this distinction may be useful in application.  

\subsection{Mixers}\label{sec:mixers}

When mapping a problem 
to quantum hardware, 
the inclusion of 
penalty terms can dramatically increase the resources required to implement a given algorithm, in terms of both the weights and density relative to the cost function of the penalty operator terms.
    
Additionally, in practice penalty-based methods do not prevent a finite or possibly significant probability of invalid and/or infeasible states, which as mentioned 
may dramatically increase the number of algorithm repetitions and hence overall 
time required to obtain a satisfactory problem solution. 

An alternative approach is to design 
quantum operations and initial states so as to automatically restrict algorithm dynamics to the subspace of feasible 
states, such that the need for penalty terms is avoided altogether. Such an approach has been developed for generalizations of AQO \cite{hen2016driver,hen2016quantum} and QAOA \cite{hadfield2017quantum,hadfield19_qaoa,ruan2020constr,bartschi20_permut,wang20_xymixers,fuchs21_maxkcut,bako2022funcqaoa}. In the context of QAOA, 
whether such an approach leads to 
fewer quantum resources than a penalty-based approach, for comparable levels of algorithm performance, 
should be analyzed on a case-by-case basis~\cite{hadfield19_qaoa,wang20_xymixers}, and so we leave this question for future work. In this section we 
address the design of mixers that 
strictly 
preserve hard constraints, 
as well as novel \textit{approximate mixer} variants that 
tolerate some manageable degree of leakage. The approximate mixers we construct require fewer quantum resources than their exact counterparts in some cases and appear particularly suitable for applications where we are willing to trade reduced circuit depth for increased classical repetitions, for example small-scale near-term experiments, though this behaviour is not generic.

For QAOA and related quantum gate model approaches  it is 
often desirable to define mixers in terms of simpler, reusable components. Following~\cite{hadfield19_qaoa}, a \textit{full mixer} $U_{\rm{M}}=U_{\rm{M}}(\beta)$ may be constructed as an ordered product of a set of 
\textit{partial mixers} $U_{M,v}=U_{M,v}(\beta)$ such that
\begin{equation} \label{eq:fullmixer}
U_{\rm{M}} = \prod_{v} U_{M,v},
\end{equation}
where ideally each $U_{M,v}$ 
is a local operator and can be implemented easily or at least efficiently. 
Importantly, the 
partial mixers do not mutually commute in general such that different orderings of the product \eqref{eq:fullmixer} can produce different full mixers. 
Here by local we mean that each $U_{M,v}$ acts nontrivially on a bounded set of qudits, 
such that in particular partial mixers acting on disjoint sets of 
qudits can be implemented in parallel. 
Clearly, if each partial mixer preserves feasibility, then so does $U_M$. 
Hence if such a mixer is utilized in QAOA along with a feasible initial state, the algorithm is guaranteed to output feasible approximate solutions only.

Partial mixers may be expressed as quantum circuits, 
as exponentials of Hermitian generators built using DQIR primitive, or as 
DQIR local 
primitives themselves 
(e.g., $\m G_\a$).
In some cases it is useful to 
further decompose into lower levels of 
partial mixers or other basic operators
%
$U_{M,v} = \prod_{w} U_{M,v,w}$
%
and so on, as desired. 
To this end it is useful to have a template of partial mixer designs applicable to different problem classes (\textit{i.e.}, domains) and types of hard constraints, with different suitability for different hardware architectures. We consider the construction of basic partial mixing designs below and in Section~\ref{sec:gdm}. 

We categorize full or partial mixers based on two primary characteristics. 
First, as mentioned, mixers may be either strict or approximate, depending on whether they allow any leakage or not; we elaborate on the approximate case below.  
designed with respect to the 
feasible subspace, 
or 
hardware-logical-level mixers 
that may be designed to 
preserve validity; the first type can be defined independently of the encoding choice, while the latter cannot. 
%
%

Given a set of hard constraints, a strict mixer should always take feasible quantum states to other feasible quantum states, as well as explore (in some sense) a sufficiently large portion of the feasible subspace. As previously identified in \cite[Sec. 3.1]{hadfield19_qaoa}, this motivates the following general design criteria for construction suitable mixing operators, in particular as products of partial mixers.

\begin{criterion}\label{crit:fullmixers}
Desired criteria for full mixers $U_M( \beta)$.
\begin{enumerate}[label=\alph*.]
\item (Feasibility) For all values of 
$\beta$ the mixer should preserve the feasible subspace, i.e., not result in any leakage when acting on a feasible quantum state. (This criteria may be relaxed if penalties or post-selection are introduced.)
\item (Reachability) For all pairs of basis states $\mathbf x \neq \mathbf y$ in the feasible subspace there must be non-zero transition amplitude overlap $\la x| (U_M(
\beta))^r |y \ra$ for some 
$\beta$ and positive integer $r$. 
\end{enumerate}
\end{criterion}
Criteria 1(a) ensures that when such a mixer is used in QAOA, only feasible (approximate) solutions will be returned. Given a set of partial mixers satisfying 1(a), any mixer formed from taking products will also preserve feasibility. Criteria 1(b) allows flexibility in terms of trading off the circuit depth per mixing layer (and hence overall number of QAOA layers given fixed resources) with the degree of mixing. 
In Section \ref{sec:gdm} we introduce techniques for the automated design of mixers satisfying Design Criteria \ref{crit:fullmixers} using graph-theoretic approaches. The design criteria is easily extended to accommodate more general cases such as multi-parameter mixers $U_M(\pmb \beta)$.

\subsubsection{Approximate mixers}
\label{sec:inexact-trott}



Here we consider generalized mixing 
operators which may allow some degree of leakage.

We call a mixer \textbf{strict} if it preserves the 
feasible subspace as in design criteria 1(a), \textit{i.e.} if $L(U_M(\beta),\ket{x})=0$ for all $|\mathbf x\ra \in |\mathbf x\ra_{\rm{feasible}}$ and all $\beta$; otherwise we use the term \textbf{approximate} mixer. If they can be implemented 
with low or moderate cost, strict mixers are preferred because they search only the valid and feasible problem space. 
However, 
there may be cases where a strict mixer 
is relatively expensive, or even when an efficient strict mixer construction is not known. In such cases an approximate mixer may be used, for which there may be a finite leakage $L>0$ per mixing stage, as well as leakage at the end of the algorithm. 
We 
elaborate on 
some additional motivating example applications for 
approximate mixers in Section~\ref{sec:pen_exch}, in particular that in some cases they may be further combined with 
penalty terms 
to reduce or eliminate leakage.

A naive implementation of an approximate mixer $\tilde U_M$ is to first define a Hermitian generator $G_M$ whose \textit{exact} exponential $U_M$ would satisfy Design Criteria \ref{crit:fullmixers} and produce zero leakage, but 
which 
cannot be 
easily implemented on a given quantum computer.  
If $G_M$ is decomposed as
\begin{equation}\label{eq:gen_decomp}
G_M = \sum_v G_{M,v}
\end{equation}
such that each $G_{M,v}$ may be exponentiated exactly, then a full approximate mixer may for example be constructed as a product formula.
When using qubits, an appropriate choice is for each $G_{M,v}$ to be a Pauli string, as quantum circuits for their exact exponentials are 
easily implemented with basic quantum gates \cite{mikeike11}. For some parameter values, or on some initial states, the leakage may be relatively small or manageable. 
For example, 
a first-order Trotter step 
corresponds to 
\begin{equation} \label{eq:trottermixer}
\tilde U_M = \prod_v e^{ -i \theta G_{M,v} } \approx U_M .
\end{equation}
which implies that Trotter error and hence leakage can be bounded as a function of $\theta$.
%
Within the same or between different mixing layers feasibility-violating transitions can cancel to some degree to have a less detrimental effect than worst-case bounds would indicate. 
%
If one wishes to reduce or control the leakage they can replace 
$\tilde U_M(\theta)$ with 
$(\tilde U_M(\theta_v/r))^r$ for roughly $r$ times the circuit cost.
Similar considerations apply to products of partial mixers derived from higher-order Suzuki-Trotter 
approximations~\cite{suzuki1985}. 
More general quantum Hamiltonian simulation algorithms may also be applied to implement $U_M$, however these approaches often appear to be beyond near-term capabilities 
and don't typically result in a products of partial mixers in the same way. 
A general takeaway is that one can often trade-off increased circuit depth with tighter leakage guarantees, when desired.
%
Importantly, 
we show in Section \ref{sec:results} that when constructing 
single-variable mixers, the naive Trotter approach \eqref{eq:trottermixer} for approximate mixers can nevertheless lead to larger circuit depths than the strict mixers designed in Section~\ref{sec:gdm_intro} of this work.

Finally, we 
remark that approximate mixers may be especially appropriate in the near-term setting, where circuit depths are limited and problem sizes not too large such that one may potentially tolerate a 
more significant decrease to the probability of success due to leakage than in the asymptotically large setting.

%

\subsubsection{Single-variable mixers}\label{sec:single-var-mixer}


Single-variable mixers (
i.e., single-qudit mixers \cite{hadfield19_qaoa}) operate on one variable, mixing only the $d=d_\alpha$ valid values of the space. We define a DQIR-level generator for such a mixer called a \textit{shift partial mixer Hamiltonian} (also called a fully-connected mixer \cite{hadfield19_qaoa}),
\begin{equation}\label{eq:shift_ham}
\begin{split}
G^{shift}_{\alpha} &= \sum_{k=1}^{d-1} \m T_\a^{(k \leftrightarrow k\text{-}1)}  \\
&= \sum_{k=1}^{d-1}  ( | k \ra\la k-1 |_\a + | k-1 \ra\la k |_\a )
\end{split}
\end{equation}
leading to a full mixer generator over all variables $G_M = \sum_{\alpha} G^{shift}_\alpha$. If the term $\tran{0}{d-1}{\a}$ is added to equation \eqref{eq:shift_ham}, the operator is called the single-qudit ring mixer Hamiltonian \cite{hadfield19_qaoa}.
As discussed in the previous section, a single-variable \textit{approximate} mixer may be implemented for encoded variables 
after further decomposing $G_\a^{shift}$ into Pauli strings which can be exponentiated exactly. 

We now turn to special cases of \textit{strict} mixers for specific qubit encodings. The more general design of mixers for arbitrary $d$ and arbitrary encodings are considered in Section \ref{sec:gdm}. 
%
%
For the unary (one-hot) encoding, we require that for pairs of qubits the two encoded states $|k\ra \mapsto \ket{0}\ket{1}$ and $|l\ra \mapsto \ket{1}\ket{0}$ be mixed,  
while $\ket{0}\ket{0}$ (which corresponds to the other encoded values) is invariant, and no leakage to $\ket{0}\ket{0}$ or $\ket{1}\ket{1}$ can occur, 
meaning that the two-qubit unary partial mixer may
have pattern
\begin{equation}\label{eq:Agate}
A_\phi^{(k,l)} = 
\begin{pmatrix}
 * & 0 & 0 & 0  \\
 0 & * & * & 0  \\
 0 & * & * & 0  \\
 0 & 0 & 0 & * 
\end{pmatrix}.
\end{equation}
A possible \textit{full} mixer in the unary encoding may thus be defined as $\prod_\a^M(\prod_{k=1}^{d-1})A_{\phi}^{k\text{-}1,k}$. Notably, these gates can be applied in parallel on qubits $\{(0,1),(2,3),\cdots\}$, followed by $\{(1,2),(3,4),\cdots\}$, meaning that the depth of this single-variable unary mixer is independent of both the number of discrete variables and the problem size. 


When using compact codes (Gray and SB), in the special case for which $d$ is a power of 2, all available encoded quantum states are valid. Therefore 
a minimal depth choice is the \textit{simple binary mixer} \cite{hadfield19_qaoa} (sometimes called the transverse-field mixer before an encoding is specified),
\begin{equation}\label{eq:sbm}
U_{\rm{M}}^{\rm{SBM}} = \bigotimes^{N_q}_j R^{X}_j(\theta_j)
\end{equation}
where $R^{X}_j=\exp(-i \theta X_j )$ is the Pauli $X$ rotation gate on qubit $j$ and $\theta_j$ are mixing parameters. 
This circuit, with a depth of only 1, is significantly shorter than in the unary (one-hot) case, where it is not possible to construct a circuit of single-qubit rotations that always preserves feasibility. We emphasize that the simple binary mixer is strict (produces no leakage) \textit{only} in the special case of $d$ being a power of 2; other cases are considered in Section \ref{sec:sgl-gdm}.

Recalling \eqref{eq:sum_tens_prod}, multi-$d$-variate mixers may be constructed by combining single-variable DQIR primitives in a similar manner as the single variable mixer case described here. 

\subsubsection{Permutation mixers}\label{sec:perm-mixer}

%

We next consider partial permutation mixers (PPMs), used for permutation problems like scheduling and routing. 
A DQIR basis state 
is valid if each  
object (for example, each city) appears exactly once; 
hence for permutations we use \emph{variable} to refer to the integer values $\pi(i)$ 
as in Eq.~\eqref{eq:x_permut}. 
%
For exploring the space of all permutations it suffices to consider PPMs that operate on two variables, from which full mixers satisfying Design Criteria~\ref{crit:fullmixers} may be built, 
and so we focus on this case. 

We introduce the following design criteria 
specific to problems over permutations. 
Criteria \ref{crit:2varppm}(a) and \ref{crit:2varppm}(b) are directly related to Criteria \ref{crit:fullmixers}(a) while \ref{crit:2varppm}(c) is directly related to \ref{crit:fullmixers}(b).

\begin{criterion}\label{crit:2varppm}
Criteria for designing a DQIR two-variable partial permutation mixer $U^{\rm{PPM}}(\beta)$. 
\begin{enumerate}[label=\alph*.]
\item 
For any pair of variables in an $M$-object permutation, 
the possible 
two-variable 
configurations $|k,l\ra$ are those for which 
$k,l \in \mathbb Z_M$ 
and $k \neq l$. 
No elements in $U^{\rm{PPM}}(\beta)$ between 
such states and necessarily infeasible states (those for which $k=l$, $k\notin \mathbb Z_M$, or $l \notin \mathbb Z_M$) are allowed.
(This criterion may be relaxed if some leakage is allowed.) 
%
\item The only allowable non-zero off-diagonal elements involving feasible 
states are those for 
DQIR-level operators $\tran{(k,l)}{(l,k)}{}$, \textit{i.e.} terms such as $\tran{(a,l)}{(l,k)}{}$ and $\tran{(a,b)}{(l,k)}{}$ are not allowed for $\{a,b,k,l\}$ all-unequal. (This criterion may be relaxed if some leakage is allowed.) 
\item The set of non-zero off-diagonal $U^{\rm{PPM}}(\beta)$ 
DQIR-level primitives $\tran{(k,l)}{(l,k)}{}$ that obey Criteria \ref{crit:2varppm}(a) and (b) must include all DQIR single-variable values $\{0,\cdots,d-1\}$ at least once.
(Note that this criterion is used toward ensuring Criterion~\ref{crit:fullmixers}(a) is satisfied in the case where we construct a full mixer using the same PPM across different variable pairs; more generally, this condition may be relaxed.)
%
\end{enumerate}
\end{criterion}
%

Full mixers may be constructed by combining two-variable PPMs in simple patterns, with the PPMs designed in accordance with  Criteria~\ref{crit:2varppm} in order to ensure that Criteria~\ref{crit:fullmixers} is satisfied. 
For example, one may again use a $U^{\rm{PPM}}(\beta)$ to operate on DQIR variable pairs $((0,1),(2,3),\cdots)$ followed by $((1,2),(3,4),\cdots)$, and repeating. Criteria for designing PPMs are further discussed from a graph-theoretic perspective in Section \ref{sec:ppm-gdm}.

We first consider a two-variable mixer Hamiltonian whose exact exponential meets Criteria \ref{crit:2varppm}. We define the two-variable standard partial permutation mixer (SPPM) Hamiltonian as
%
\begin{equation}
\begin{split}
H_{\a,\b}^{SPPM} = \sum_{k=1}^{d-1} \Big( \tranasymm{k}{k\text{-}1}{\a} \tranasymm{k\text{-}1}{k}{\b} \\+ \tranasymm{k}{k\text{-}1}{\a} \tranasymm{k\text{-}1}{k}{\b} \Big)  \\
= \sum_{k=1}^{d-1} |k \kb k-1|_{\a} \otimes |k-1 \kb k|_{\b} \\+ \sum_{k=1}^{d-1} |k-1 \kb k|_{\a} \otimes |k \kb k-1|_{\b}.
\end{split}
\end{equation}
%
The exact exponential of this operator would ensure that two of the same integers would never appear more than once, which in turn ensures that the state remains in the feasible space for permutation problems. 
As discussed, an approximate SPPM mixer can be derived from a Suzuki-Trotter product formula, which will (depending on the encoding) 
lead to some degree of leakage. 

In the unary encoding, it is straight-forward to define a \textit{strict} partial permutation mixer to mix state $k$ on variable $\a$ with state $l$ on variable $\b$. One may use a gate of the form $A_\phi$ of equation \eqref{eq:Agate}, where the two target qubits correspond to states $|k\ra_\a$ and $|l\ra_\b$. (These have been called ordering swap partial mixers \cite{hadfield19_qaoa}.) Implementing 
these gates for sufficiently many different pairs states will lead to a PPM that meets criterion \ref{crit:2varppm}(b).

Strict partial permutation mixers for standard binary, Gray, and block unary encodings are much less straight-forward to design, even when $d$ is a power of 2. Unlike the single-variable mixer case, there is no two-variable PPM equivalent of the simple binary mixer of equation \eqref{eq:sbm}, because such a mixer would lead to infeasible states such as $|k,k\ra$. Novel graph-theoretic strategies for designing PPMs are discussed in Section \ref{sec:ppm-gdm}.


\subsubsection{Combining mixers with penalties}
\label{sec:pen_exch}


Here we explain how in some cases it may be advantageous for algorithms such as QAOA to combine the mixers and penalty term approaches, which we refer to generally as \textit{penalty exchanging}. In this approach, we select a mixer that preserves some superset $\mathbf D' \supset \mathbf D_{\text{feas}}$ of the feasible subspace, and as needed add penalty terms to the cost function (cost Hamiltonian) to suppress transitions to strings $\mathbf y\in \mathbf D'\setminus \mathbf D_{\text{feas}}$.  If $|\mathbf D'|$ is not too much larger than $|\mathbf D_{\text{feas}}|$ it may be possible to avoid penalty terms altogether. Here, a strict mixer on $\mathbf D'$ may be approximate with respect to the target subspace $\mathbf D_{\text{feas}}$. 
In particular this approach may be applied at the level of individual variables and domains $D_\alpha$. Furthermore, in some cases it may be possible to select an exact mixer with respect to $\mathbf D'$ such that 
all or some 
measurement outcomes $\mathbf x'\in \mathbf D'\setminus \mathbf D_{\text{feas}}$ can be classically `corrected' to a feasible string $\mathbf x\in \mathbf D_{\text{feas}}$, e.g. the approach of \cite{streif2021quantum}.  

We describe some scenarios where this approach may be useful. 
For some problems it may be possible to reduce required mixer resources such as circuit depth by relaxing the individual variable domains (i.e., increasing $d$), at the expense of tolerating some degree of leakage. 

For example, imagine a 
single-variable mixer for a $d=14$ variable that requires much deeper circuits than that for $d=16$ in a given compact encoding. It may be the case that depth can be lowered by 
extending the cost function to $d=16$ while adding a variable domain penalty $F_{SS}$ for $d=14$. 
We consider such an example in Sec. \ref{sec:simple}. 
Another example is that for some hard constraints it may by hard or inefficient to construct a mixer that exactly preserves the feasible subspace. In this case we may relax the domain to one for which a suitable mixer can be efficiently implemented, and augment the cost Hamiltonian with appropriate penalties.    
A third example is that, generically, a mixer which is exact at the DQIR level may, after encoding, be compiled in an approximate way (i.e., may allow some leakage to invalid states).  

For a given problem and domain, different combinations of mixers and penalty terms may be possible. In order obtain decreased circuit depth, these must be selected such that the gains for the mixing stage are not outweighed by the added cost of implementing the penalty terms. Moreover, any gains should lead to manageable degree of leakage with respect to the reduction of probability of success. A further critical consideration beyond the scope of this article is the effect on overall performance.

For specific realizations, penalty exchange ought to be analyzed or numerically tested within a full quantum algorithm. 
Exploring suitable circuits using penalty exchanging (based  for example on memoized template circuits of fixed depths) is a task appropriate for DQIR.

\subsection{Initial states}\label{sec:init}

There are many options for the input quantum state. In practice, it is difficult to know \textit{a priori} which combination of initial state and algorithm 
is most favorable for a given problem instance and quantum device. There is a complex interplay between choice of initial state, encoding, penalties, mixers, and algorithm. For instance, if one allows for use of penalties then one is less restricted in the choice of initial state, but one may pay a substantial price in 
obtaining feasible outcomes~\cite{wang20_xymixers}. 

Here we categorize the selection of initial state in terms of several characteristics. 
Algorithms such as QAOA or other variational approaches in principle can accomodate arbitrary initial states when the link to adiabatic evolution is relaxed. 
However, in general a given target solution may be more difficult or even impossible 
to access from one initial state than from another~\cite{akshay2020reachability}, 
and further research is needed to better understand the performance and resource tradeoffs.

One may begin with either a well-defined state for which a quantum circuit is known or a quantum state that was outputted from a previous quantum algorithm (for instance, output from QAOA may be used as input for an ITE algorithm). Here we focus on the former case, where the \textit{initial} state should be easy-to-prepare relative to the cost of the circuit to follow~\cite{hadfield19_qaoa}.

An important general class of states 
consists of tensor products of single-variable states, which may be  
expressed as $\ket{\psi_0} = \otimes_\a^{M} |\phi_\a\ra$ for arbitrary single-variable states $|\phi_\a\ra$. 
The primary advantage of tensor product states is that they can be prepared in short depth.
If there are no feasibility constraints, one may choose to prepare an equal superposition of all valid states
\begin{equation}\label{eq:sup_feas}
|\psi_0\ra = \frac{1}{\sqrt{Md}} \prod_{\a}^M \left( \sum_{k=0}^{d-1} |k\ra \right)
\end{equation}
by operating on each (encoded) variable in parallel. This encapsulates the 
trivial-to-prepare 
qubit initial state $|\psi_0\ra=|+\ra^{\otimes N_q}$ as originally proposed for QAOA. 
In addition to being easy to prepare, 
equal superposition states 
are also the ground state of a mixer Hamiltonian 
that is a sum of single-variable terms.

For permutation problems 
such as scheduling and routing, whose solution is in $\{\mathbf x\}_{\rm{perm}}$, it may be 
advantageous to begin in a superposition of valid permutations
\begin{equation}
\sum_{k} c_k | \pi_k(1) \ra_1 | \pi_k(2) \ra_2 \cdots | \pi_k(M) \ra_M
\end{equation}
where $\{\pi_k\}$ is the set of all valid permutations and $c_k$ are 
complex coefficients. Previous work has shown how to prepare an equal superposition of permutations in the unary (one-hot) encoding \cite{bartschi20_permut}, with a cost scaling as $O(M^3)$ in the gate count and $O(M^2)$ in the depth. This highlights that although more clever initial states may lead to more robust numerical behavior, one may pay a higher cost in the initial state preparation. We observe that, using known conversions \cite{sawaya20_dlev} from unary to standard binary, Gray, and block unary encodings, this unary superposition state may be converted to any of the latter encodings.

Importantly, single-variable tensor product states 
subsume the special case of initializing a quantum algorithm with a classical string, that may selected randomly, or be obtained with some 
amount of instance-dependent classical preprocessing, for example using the output of a 
classical algorithm or heuristic. Such initial state subroutines may range from simple greedy algorithms to much more sophisticated approaches.  
One may 
integrate such classical solvers at the DQIR layer in the software pipeline. 
Given such a classical state $\ket{\pmb y}$, we may always transform to a quantum superposition, if desired, by applying a single mixing stage~\cite{hadfield19_qaoa}, or other suitable operator.

Other useful variations on the above initial states are possible. One may introduce classical or quantum randomness into an ensemble of initial states as a way to explore more of the solution landscape, and mitigate the possibility of a poor choice. 
For example one may implement a ``warm-starting'' procedure \cite{egger20_warm}, where output from a classical relaxation (whereby discrete variables are replaced with continuous ones) of a combinatorial problem is used to 
derive an input quantum state and appropriate mixer in a different way than described above. Numerous other initial state variations are possible including adaptive approaches. 
Generally, an initial state design should be evaluated and selected in tandem with the intended algorithm (e.g., mixer) toward achieving the best possible performance.

\section{Graph-derived mixers}\label{sec:gdm}

In this section we show a novel approach to designing suitable mixing operators using a search algorithm based on graph-theoretic design criteria. We provide numerics supporting the resource advantages of our approach in Section~\ref{sec:results}. 

The purpose of this section is to introduce a general method for creating mixers that strictly preserve the valid and feasible subspace. There have been several works that introduced their own class of such constraint-preserving mixers \cite{bartschi20_permut,wang20_xymixers,fuchs21_maxkcut}. Grover mixers \cite{bartschi20_permut} use a construction based on Grover's algorithm that yields constraint-preserving circuits that include multi-Toffoli gates; our approach provides additional flexibility and may lead to shorter gate depths, and one may choose to construct mixers from at most 2-qubit gates, which do not need to be decomposed. Another notable work \cite{fuchs21_maxkcut} constructs constraint mixers on compact (e.g. binary) encodings; our current work produces considerably shallower mixer circuits.
Finally, constructs such as XY mixers \cite{wang20_xymixers,fuchs2022constrainedonehot} are efficient mixers that apply only to one-hot encodings; in contrast, our general graph based approach may in principle be applied to any encoding and any constraint. 
Despite our graph approach offering more flexibility in gate choice as well as shorter circuit depths, there are some potential drawbacks. First, the search algorithm in our graph-derived mixers may lead to longer compile times than previous approaches. Second, there is no guarantee that an appropriate mixer is found, if the choice of gate library is not sufficiently powerful (however, were were able to design mixers for all use cases considered in this work). Hence, determining which constraint-preserving approach to take must be studied on a case-by-case basis.

\subsection{Graph representations for strict mixers}\label{sec:gdm_intro}

Here we introduce a novel approach in which we represent partial mixers as graphs derived from their action on computational basis states. We will show how such a construct facilitates the automated design of strict partial and full mixers that meet Design Criteria \ref{crit:fullmixers}.  
These partial mixers are encoding-dependent and must be designed separately for each encoding. 
In general the size of the mixer graph grows exponentially with the number of problem variables~$M$, in the worst-case. 
However, there are many problems (including permutation problems and coloring problems considered later in this work) for which a full mixer across all variables may be 
generated from applying the same 
one- or two-variable \textit{partial} mixers across different variables, for which the size of the corresponding graph grows with $d$ (which is often a constant) rather than $M$.

Consider a partial mixer unitary $U_{PM}$ operating on $L\leq M$ quantum discrete variables (i.e., $U_{PM}$ 
is designed in terms of its action on a Hilbert space of size 
$\m D_{PM} \leq d_1 \times \cdots \times d_L$ where $d_\a$ is the cardinality of variable $\a$). We define the \textbf{partial mixer graph} (PMG) as  
\begin{equation}
\Omega(U_{PM}) = ( V, \{ uv | U_{PM}[u,v] \neq 0 \} )
\end{equation}
where the set of vertices $V$ corresponds to the $\m D_{PM}$ quantum basis states and $U^{PM}[u,v]$ corresponds to the coefficient of 
$|u \kb v|$, \textit{i.e.} $U^{PM}[u,v]\propto \textrm{trace}(U^{PM}|u \kb v|)$. 
The graph $\Omega$ is undirected, as 
for simplicity in this work we restrict ourselves to the class of unitaries for which $U^{PM}[u,v] \neq 0$ if and only if $U^{PM}[v,u] \neq 0$. Our approach similarly extends to the more general directed case.

Consider for example 
a controlled-$R_Y$ operator with the first and second qubit as target and off-control, respectively, in the context of its action on some (encoded) 3-qubit state (\textit{i.e.}, the middle green box in the bottom left of Figure \ref
{fig:hard-sgl-mixer}).  
This gate has the following sparsity pattern and the resulting partial mixer graph contains two edges:

\begin{equation}
\begin{split}
U \sim 
\begin{pmatrix}
 * & 0 & 0 & 0 & * & 0 & 0 & 0 \\
 0 & * & 0 & 0 & 0 & * & 0 & 0 \\
 0 & 0 & 1 & 0 & 0 & 0 & 0 & 0 \\
 0 & 0 & 0 & 1 & 0 & 0 & 0 & 0 \\
 * & 0 & 0 & 0 & * & 0 & 0 & 0 \\
 0 & * & 0 & 0 & 0 & * & 0 & 0 \\
 0 & 0 & 0 & 0 & 0 & 0 & 1 & 0 \\
 0 & 0 & 0 & 0 & 0 & 0 & 0 & 1 \\
\end{pmatrix} \\ \\
\;\mapsto \; \Omega(U) = \{ (0,4); (1,5)  \},
\end{split}
\end{equation}
%
where 
here we have used $\{\ket{000},$ $\ket{001},$ $\ket{010},$ $\ket{011},\cdots\}$ $\mapsto$ $\{0,1,2,3,\cdots\}$. This construct is useful because we may now redefine mixer design as a graph-theoretic problem, which will often be both conceptually 
and computationally simpler 
than using 
the usual operator matrix representations.



For a fixed set of problem hard constraints, we define $S_G$ and $S_B$, respectively, as the induced sets (subspaces) of \textit{locally} good (feasible) 
and 
bad (infeasible or invalid) states, defined 
with respect to the $L$ variables under consideration. 
This approach facilitates easy reuse of the resulting local partial mixers independently of the global problem structure. 
It is vital to emphasize that a locally good state in $S_G$ 
may or may not 
correspond to a feasible \textit{global} state. 
Consider for example a problem over permutations $\{\pmb x\}_{\text{perm}}$ where we wish to design mixers with $L=2$. For any two variables, $S_G$ contains all states for which the two variables are assigned different integers. However, if in the global variable space a different variable is assigned the same integer as one of these two local states, then the global basis is \textit{not} a feasible state (even though the two-variable local state is feasible).  
%
%

Using $S_G$ and $S_B$ we restate our 
mixer design criteria in graph-theoretic terms below. In particular, the 
following condition must be met for any mixer design: \textit{Edges between nodes in $S_G$ and nodes in $S_B$ are not allowed.} Additional design criteria are often necessary in considering specific domains or problem classes.

This graph-based 
perspective can be 
particularly useful, via the following approach, for constructing low-depth mixers satisfying the relevant design criteria.  
Observe that in order to 
design local partial mixers (\textit{i.e.} local unitaries) that preserve the valid subspace, one may first construct a library of short quantum circuits, for which it suffices to store only the graph $\O(U)$ for each circuit. Then, given a set of hard constraints (on $M$ variables), an automated search over this library of (local) graphs, along with a set of graph theoretic design criteria implementable in a compiler, 
allows one to design circuits for, in principle, arbitrary sets of constraints (see Figure \ref{fig:hard-sgl-mixer}). This search may be done during compilation, or before. We call these \textbf{graph-derived partial mixers} (GDPM). As mentioned, in a number of settings, such as when partial mixers acting on bounded sets of qubits suffice, this process is guaranteed to be efficient.  Moreover, these partial mixers can be easily reused across different problems or instances. 

We remark that a future possibility for dealing with more sophisticated hard constraints is to use ancilla-based mixer designs, which have been proposed previously \cite{hadfield19_qaoa}. For example, one may compute arithmetic information to store in ancilla qubits, before using 
this register as control qubits to perform a 
subsequent mixing operation only if it is guaranteed to 
preserve the feasible space. Such procedures 
are relatively straightforward to design and implement in DQIR, in an encoding-independent way. For example, this may be a workable approach for 
dealing with hard constraints in integer linear programming, though we do not explore this direction in detail here. 
Generally, there exist a variety of open questions related to the design of more effective mixers~\cite{hadfield19_qaoa}; 
we 
remark on several important directions here in Section~\ref{sec:concl}.

\subsection{Single-variable GDPMs}\label{sec:sgl-gdm}

\begin{figure}
    \centering
    \includegraphics[width=.95\linewidth]{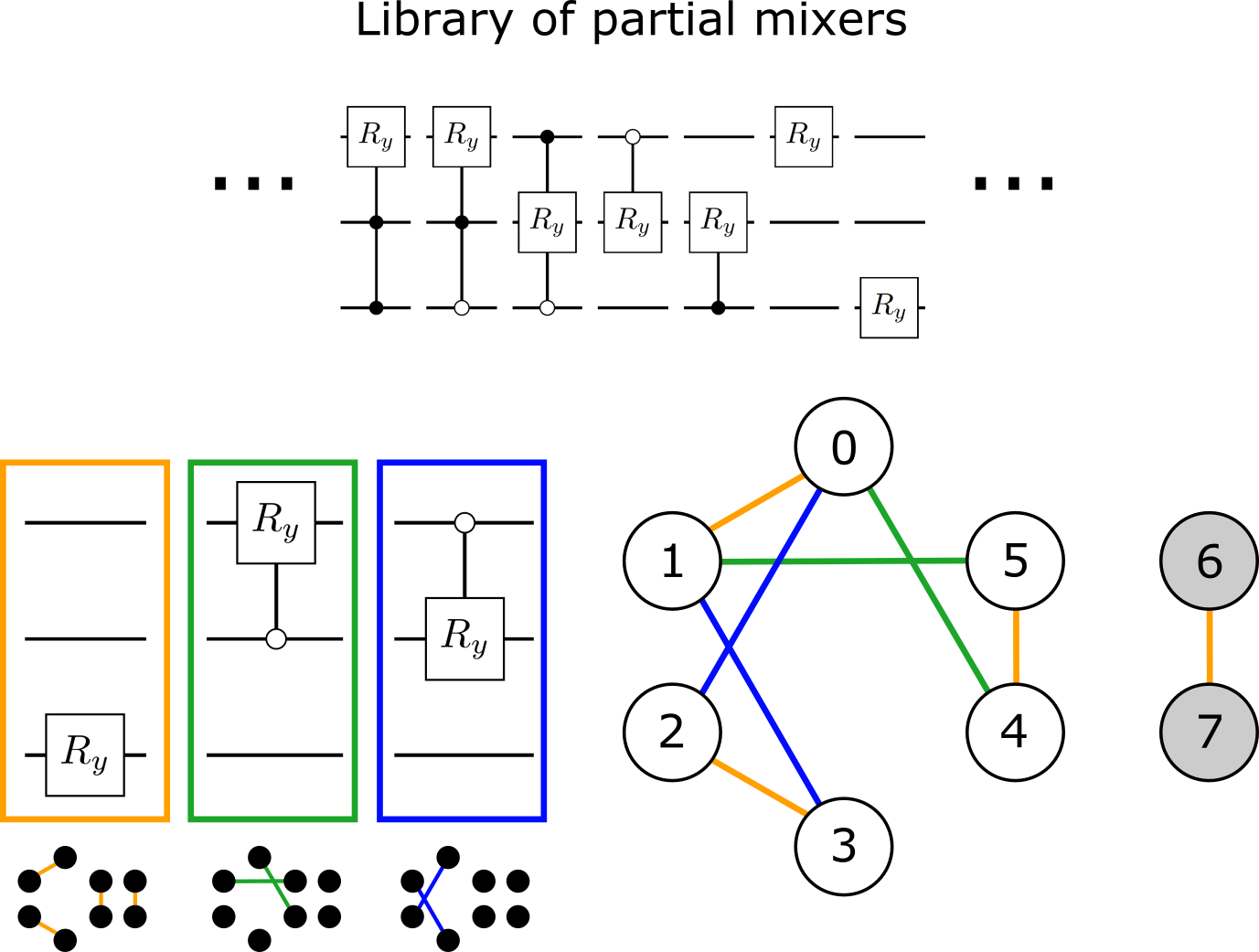}
    \caption{A schematic of the mixer design procedure for hard constraints. First one compiles a library of partial mixers. Each partial mixer yields a graph corresponding to which states are connected by off-diagonal terms. The example in the image is for single-variable GDPMs, which may be used for example in coloring problems. Design Criteria \ref{crit:sglvar_gdpm} dictates that the union of the partial mixers' graphs must yield a graph for which all valid states form a connected graph, and there must be no edges between valid and invalid states. The example shown is a single-variable mixer design for the SB encoding of $d=6$. Nodes $|0\ra=|000\ra$ through $|5\ra=|101\ra$ form a connected graph, from which invalid nodes $|6\ra=|110\ra$ and $|7\ra=|111\ra$ are disconnected.}
    \label{fig:hard-sgl-mixer}
\end{figure}

The simple 
mixer $U_{\rm M}^{\rm{SBM}}$ of equation \eqref{eq:sbm} is a single-variable mixer for compact codes that may be used especially when $\log_2 d$ is an integer. But designing strict mixers for compact codes is more difficult when $d$ is \textit{not} a power of 2, and it is 
with this case in mind that we develop novel classes of mixers. Here we introduce a general approach for designing strict mixers 
satisfying our criteria, based on the graph-theoretic approach introduced in the previous section. 
%
We first define two specialized criteria \ref{crit:sglvar_gdpm} 
for designing a compact single-variable mixer. Criteria \ref{crit:sglvar_gdpm}(a) and (b) are related to Criteria \ref{crit:fullmixers}(a) and (b), respectively.
\begin{criterion}\label{crit:sglvar_gdpm}
Criteria for designing a single-variable GDPM from simpler partial mixers for a given $d$ and encoding choice. 
\begin{enumerate}[label=\alph*.]
    \item Edges between sets $S_G$ and $S_B$ are not allowed. 
    \item The union of the PMG 
    subgraphs restricted to $S_G$ must yield a connected graph (\textit{e.g.} there must be some path between any two arbitrary feasible states). 
\end{enumerate}
\end{criterion}

\begin{figure*}[t]
    \centering
    \includegraphics[width=.9\linewidth]{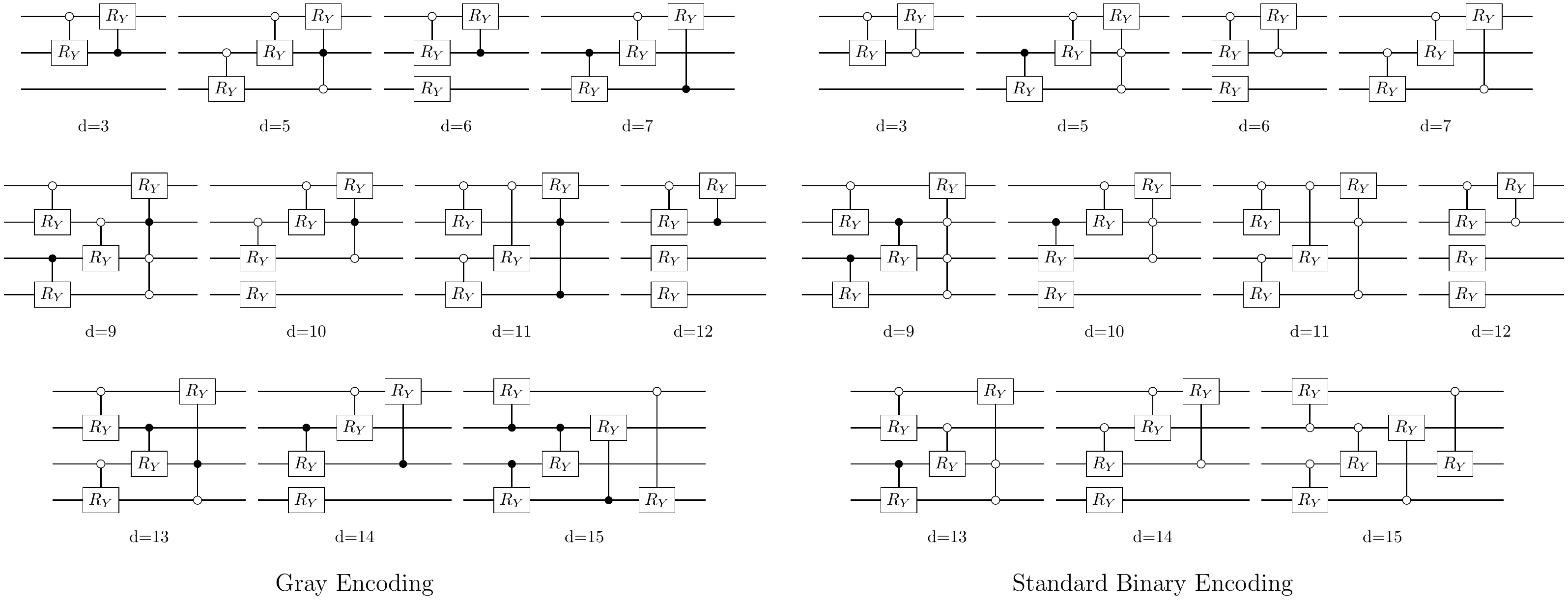}
    \caption{
    Computationally designed graph-derived partial mixers (GDPMs) for single-variable mixers that meet Design Criteria \ref{crit:sglvar_gdpm}, for SB and Gray encodings. Note that these encodings require $\ceil{\log_2 d}$ qubits. This is just one possible set of GDPMs, determined by our heuristic algorithm outlined in Algorithm Sketch \ref{algo:sgl-var-mix} and an arbitrary choice of unitary library. The low-depth $U_{\rm M}^{\rm{SBM}}$ of equation \eqref{eq:sbm} may be used when $d=2,4,8,16,$ etc.}
    \label{fig:mixer_circuits}
\end{figure*}

These criteria lead 
to the following 
algorithm for 
designing strict single-variable mixers for 
arbitrary 
$d$ 
and encoding choice. A library of basic (parameterized) unitaries must be 
input, from which the algorithm finds an ordered product of 
library elements that meet Criteria \ref{crit:sglvar_gdpm}. 
The library must be expressive enough to meet the design criteria, and may consist, for instance, of previously determined partial mixer designs or particular favorable operations on a given hardware device.  The design of a single-variable GDPM is shown schematically in Figure \ref{fig:hard-sgl-mixer}, where three primitive circuit elements are combined to produce a mixer for $d=6$ in the SB encoding.
The following algorithm sketch shows how a GDPM may be found via converting a library of parametric circuit unitaries into graph representations. While here we consider the problem of merely finding a suitable mixer, it is straightforward to extend this approach to one where some designs are favored over others, for example, if each library element came with an associated implementation cost. 

%
\begin{algo}\label{algo:sgl-var-mix} 
Algorithm for the design of single-variable GDPMs. 
\begin{enumerate}
    \item Input or construct a suitable library of parameterized unitaries $\{U_{j}^{\rm{PM}}\}$, for which each member is efficient to implement. The 
    library 
    may be expanded to include additional operators if the algorithm fails to produce a mixer that meets the design constraints.
    \item For each parameterized unitary determine $\O_{j}^{\rm{PM}} \equiv \O(U_{j}^{\rm{PM}})$, the partial mixer graph (PMG). Call this library of graphs $\mathfrak L$. For the remainder of the algorithm we use a subset of this library of graphs, as opposed to the original library of circuit unitaries.
    \item For each graph in $\mathfrak L$, discard any $\O_{j}^{\rm{PM}}$ that contains edges between $S_G$ and $S_B$. 
    The remaining members of $\mathfrak L$ cannot cause leakage.
    \item Replace each remaining member of $\mathfrak L$ with its induced subgraph $\O^{\rm{PM}}_j[S_G]$ (\textit{i.e.} the graph for which $S_B$ nodes are removed). 
    This step improves algorithm efficiency because the bad states in $S_B$ need not be stored or processed.
    \item Initialize set $\mathfrak M$ and add all members of $\mathfrak L$ to $\mathfrak M$. (In the subsequent steps, the goal is to find a union of graphs that forms a connected graph. This is equivalent to finding a union of graphs that yields a single graph component. For example, the union of the three graphs in Figure \ref{fig:hard-sgl-mixer} yields a single graph component over $S_G$.)
    \item \label{itm:beginloop} For each graph in $\mathfrak M$ find the number of graph components. 
    Remove all members of $\mathfrak M$ that do not have exactly the minimum number of graph components. (This step just ``pares down'' the number of candidate graph sets, in order to reduce the cost later on. The step is not overly stringent, as the members of $\mathfrak L$ are used again in the next step, i.e. the unitaries of the original library are continuously being recycled.) 
    \item \label{itm:endloop} For every pair of graphs $(\O\supr{\mathfrak L}_i,\O\supr{\mathfrak M}_j)$ with $\O\supr{\mathfrak L}_i \in \mathfrak L$ and $\O\supr{\mathfrak M}_i \in \mathfrak M$, add the graph union $\O_{ij}=\O\supr{\mathfrak L}_i \cup \O\supr{\mathfrak M}_j$ to the set $\mathfrak M$. (Stated differently, this step combines each graph in $\mathfrak M$ with each graph in the original library $\mathfrak L$.)
    \item Repeat steps \ref{itm:beginloop} and \ref{itm:endloop} until some graph in $\mathfrak M$ yields a single graph component. A single graph component signifies that there is some path connecting all nodes in $S_G$. Hence the unitaries that were implicitly used to compose the single-component graph may be combined in product to form  a proper single-particle GDPM, as in  Figure \ref{fig:hard-sgl-mixer}. (These unitaries may be implemented in any order; for instance the shortest-depth ordering may be chosen.)
\end{enumerate}
\end{algo}

Additional steps may be taken to improve algorithm efficiency, for example by removing from $\m S$ the PMGs that correspond to circuits with larger depths. In this work, for the design of single-variable GDPMs, we chose to use multi-controlled Pauli $Y$ rotations where the rotation qubit may be on any qubit and control qubits may be on or off. 
This is just a small subset of the possible unitaries one could consider using in a library.

The results of our computationally designed GDPMs are shown in Figure \ref{fig:mixer_circuits}, for Gray and SB encodings up to $d=15$. Similar constructions follow for arbitrary $d$, as desired. These mixer circuits are a 
primary contribution of this paper, as practitioners may directly use them 
in QAOA, particularly in cases of limited quantum resources. Decompositions of the circuit library into one- and two-qubit gates is discussed further in Section \ref{sec:decomp}. In Section \ref{sec:results} we will demonstrate that the decomposed circuit depths of these \textit{strict} mixers are substantially shorter even than the \textit{approximate} mixers that result from standard approximations to the exponential of equation \eqref{eq:shift_ham}.

GDPMs for the block unary (BU) encodings may be designed by using a compact GDPM within each $g$-qubit block, and then connecting blocks with separate unitaries. We observe briefly that for the BU$^{Gray}_{g=3}$ encoding, two blocks of two qubits each may be connected using a four-qubit gate, the doubly controlled $A_\phi$ gate. Decomposed depths of BU$^{Gray}_{g=3}$ GDPM circuits are also given in Section \ref{sec:results} as well. 

\subsection{Permutation GDPMs}\label{sec:ppm-gdm}

\begin{figure}
    \centering
    \includegraphics[width=.95\linewidth]{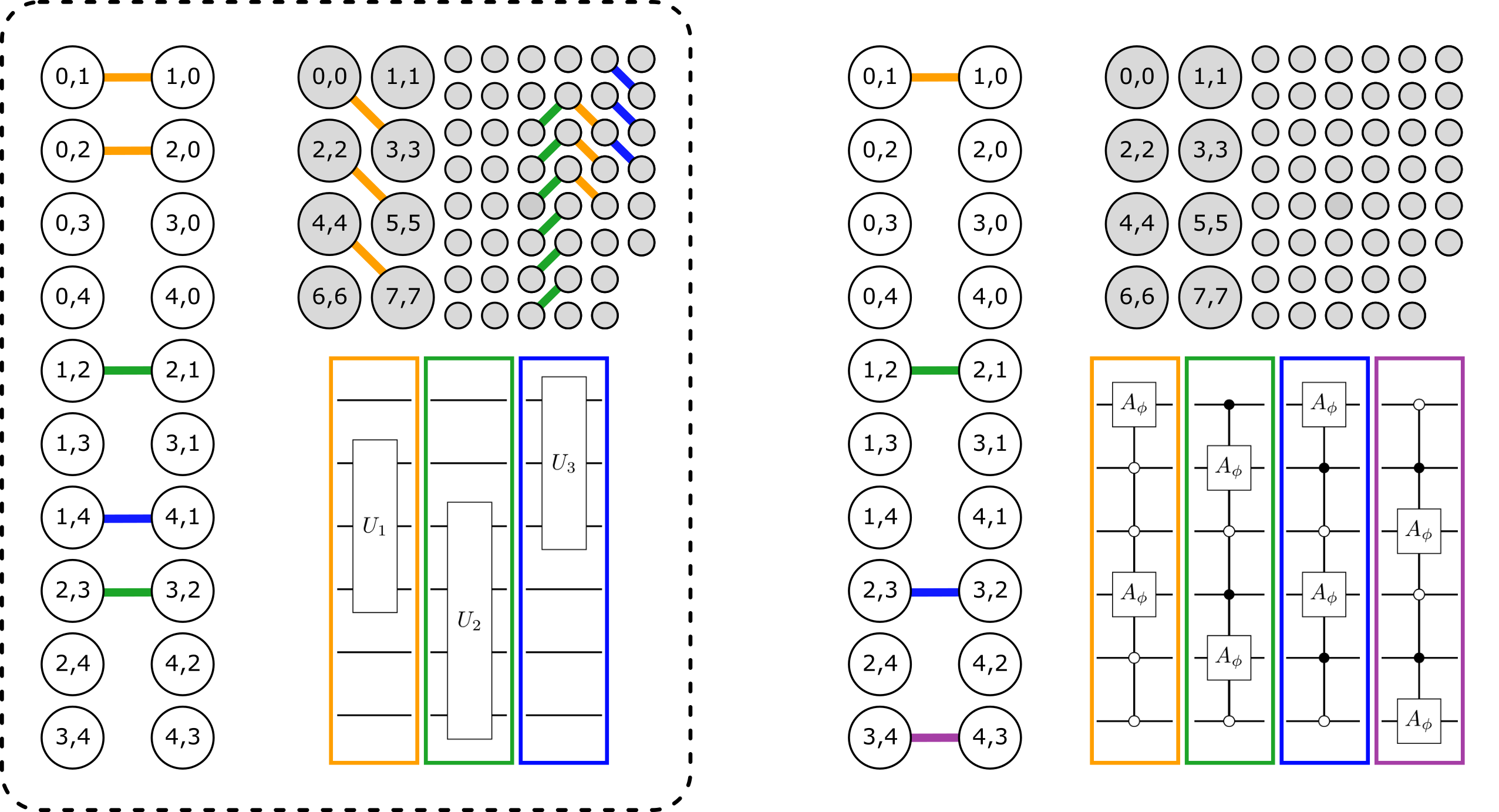}
    \caption{Design of partial permutation mixers. The graph theoretic conditions for a PPM design are considerably different from those for single-variable mixer design. The three conditions for a PPM are that (a) the set of good (white) and bad (gray) states must be disjoint, (b) the only edges allowed between the good states are those between $|i,j\ra$ and $|j,i\ra$ (horizontal edges), and (c) $\m Q$ defined in Criteria \ref{crit:ppm_gdpm}(c) must be a connected graph. 
    \textbf{Left}: A schematic of a valid \textit{hypothetical} set of partial mixers, for $d=5$. \textbf{Right}: a set of four unitaries that form a valid PPM for the Gray code and $d=5$.
    }
    \label{fig:hard-ppm-mixer}
\end{figure}

\begin{figure}[h]
    \centering
    \includegraphics[width=.95\linewidth]{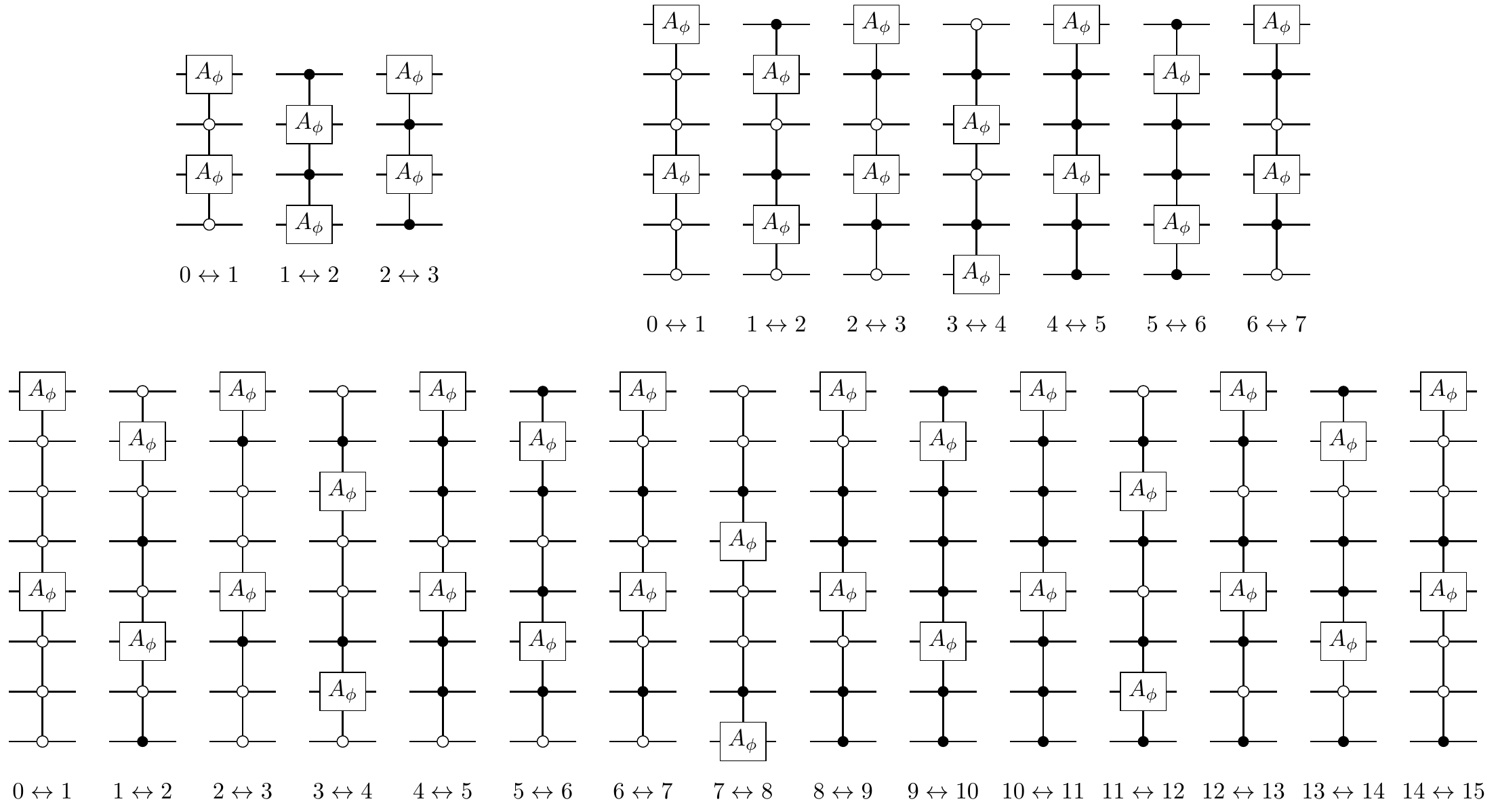}
    \caption{Two-variable partial permutation mixers for the Gray code, designed based on the criteria of section \ref{sec:ppm-gdm}. These may be easily modified for the SB encoding by using Gray-to-SB conversion circuits \cite{Shukla15_sb2gr,sawaya20_dlev} for each of the two variables.}
    \label{fig:ppm_circuits}
\end{figure}

We next introduce two-variable GDPMs for permutation problems like scheduling and routing. 
Considering again two permutation variables as in \eqref{eq:x_permut}, 
we define the set of good states $S_G$ as all pairs $|k,l\ra$ for which $k \neq l$ and $k,l < d$; all other states belong in $S_B$. The graph-theoretic conditions for designing partial permutation mixers are then quite distinct from the 
previous case:
\begin{criterion}\label{crit:ppm_gdpm}
Criteria for the design of two-variable GDPMs for permutation problems. (These are equivalent to Criteria \ref{crit:2varppm}, reformulated for the graph perspective.)  
\begin{enumerate}
[label=\alph*.]
    \item Edges between sets $S_G$ and $S_B$ are not allowed. 
    \item Within $S_G$, edges may exist \textit{only} between DQIR-space two-variable pairs $|k,l\ra$ and $|l,k\ra$.
    \item Define a graph $\m Q$ with $d$ nodes labeled $\{0,\cdots,d-1\}$, where an edge between nodes $k$ and $l$ exists iff there is an edge between $|k,l\ra$ and $|l,k\ra$. $\m Q$ must be a connected graph.
\end{enumerate}
\end{criterion}
These criteria are shown schematically on the left of Figure \ref{fig:hard-ppm-mixer}. The left-hand figure is for illustrative purposes and does not correspond to an actual mixer we construct; the right-hand figure corresponds to a 
strict mixer for the Gray code that is discussed below.

Observe that the edge placement rules are more stringent for permutation GDPMs than for single-variable GDPMs. This is because for permutations it is possible to preserve the ``local feasibibility'' condition 
(Criteria \ref{crit:ppm_gdpm}(b))
of the two-variable space while violating the feasibility of the full quantum state. Explicitly, as alluded to in Section \ref{sec:perm-mixer}, while a transformation such as $| a,k \ra\la k,l | \otimes I \otimes I \otimes I$---which does not violate Criteria \ref{crit:ppm_gdpm}(a) but does violate Criteria \ref{crit:ppm_gdpm}(b)---leads to a two-variable state on the first two variables that is in $S_G$, it still leads to an infeasible global state, for instance $|k,l,a,b,c\ra \rightarrow |a,l,a,b,c\ra$. 


%
One may define an algorithm for finding the PPMs that is 
structurally similar to Algorithm Sketch \ref{algo:sgl-var-mix}, where the number of covered local states $i$ is maximized in lieu of minimizing the graph components.
In preliminary numerical experiments, our elementary library of multi-controlled-$R_{Y}$ gates was not able to meet Criteria \ref{crit:ppm_gdpm}, as anticipated, meaning that a different library of unitaries would be needed in practice.
However, we were able to rationally design a set of PPMs that meet the correct graph-theoretic criteria, as we describe presently. 

The right side of Figure \ref{fig:hard-ppm-mixer} shows an example of our novel mixers for the Gray code with $d=5$. The gate $A_\phi$ is a rotation gate in the class of equation \eqref{eq:Agate}. The key insight behind our use of these multi-controlled $A_\phi$ is that, in the Gray code, the \textit{Hamming distance} between encoded states $|k,k+1\ra$ and $|k+1,k\ra$ is always exactly 2. Hence we ensure that the two-qubit gate $A_\phi$ mixes only the two differing bits, controlling on the bits of constant value.

The resulting GDPM circuits shown in Figure \ref{fig:ppm_circuits} may be combined for use with the Gray code, for any value of $d$ (equivalently, $M$) up to 16. For instance, for $d=10$ the first 9 gates of the bottom row of Figure \ref{fig:ppm_circuits} will together form a valid PPM. 

For compact codes generally, 
Gray code mixers may be easily converted to mixers for SB, and vice versa. One approach simply uses a circuit to convert SB to Gray \cite{Shukla15_sb2gr,sawaya20_dlev}, implements the Gray code mixer shown on the right side of Figure \ref{fig:hard-ppm-mixer}, and then converts back to SB. This adds a depth of just $2 \ceil{\log_2 d}$ CNOT gates to the overall partial mixer.

\subsection{Comments on decomposing multi-qubit gates}\label{sec:decomp}

Because the GDPMs presented here operate in the qubit space, it is worthwhile to elaborate on a few known qubit-based quantum gate decompositions that lead to shorter-depth implementations than the most commonly used decomposition might naively yield. Here we consider decompositions of the two-qubit controlled-$R_Y$ gate, the multi-controlled-$R_Y$ gate, the $A_\phi$ class of gates, and alternatives to the multi-controlled-$X$ gates like Toffoli gates. In this section, we assume a gate set consisting of CNOTs and arbitrary one-qubit rotations.

Though a general two-qubit controlled-unitary requires two CNOTs and a depth of 5, a more constrained controlled-$R$ gate may be implemented with one CNOT and two single-qubit gates for a depth of 3 \cite{barenco1995}. Such a depth-3 circuit appears to be the optimal choice for the two-qubit gates in the above-mentioned unitary library (Figure \ref{fig:hard-sgl-mixer}).
Regarding gates on more than two qubits, decompositions for multi-controlled-$R_Y$ circuits have been shown to be implementable in a depth of $2^{N_q} \sim 2d$, which is often shorter depth than general multi-controlled-$U$ unitaries \cite{slepoy06_sandia}.

As previously mentioned, a gate $A_\phi$ with the sparsity pattern of equation \eqref{eq:Agate} can be implemented in a depth of just 5 \cite{gard20_symm}, a reduction from the depth-9 gate resulting from a more naive approach of exponentiation of $XX+YY$ with the standard staircase circuit construction \cite{mikeike11}.

Some comments are merited regarding decompositions of higher-order gates as well. Depending on the methods used, the iterative process of decomposing a multi-controlled unitary often leads to intermediate multi-controlled-$X$ gates. If ancilla qubits are available, such a multi-controlled-$X$ gate can be decomposed into $O(N_q)$ Toffoli gates \cite{barenco1995}. This linear scaling in the number of qubits corresponds to a logarithmic scaling $O(\log_2 d)$ in the size of a variable's domain. Despite this seemingly favorable scaling, decomposing one Toffoli gate yields a circuit depth of 12 \cite{shende2008_toffdecomp}, leading to a considerable overall cost. 

However, here we briefly introduce an alternative strategy to effectively reduce the depth. If one relaxes the condition that all phases in the Toffoli be $+1$, then one may implement unitaries that are ``congruent modulo phase shift'' to the Toffoli gate, such as
\begin{equation}
U_{\rm{pT}} = 
\begin{pmatrix}
 1 & 0 & 0 & 0 & 0 & 0 & 0 & 0  \\
 0 & 1 & 0 & 0 & 0 & 0 & 0 & 0  \\
 0 & 0 & 1 & 0 & 0 & 0 & 0 & 0  \\
 0 & 0 & 0 & 1 & 0 & 0 & 0 & 0  \\
 0 & 0 & 0 & 0 & -1 & 0 & 0 & 0 \\
 0 & 0 & 0 & 0 & 0 & 1 & 0 & 0  \\
 0 & 0 & 0 & 0 & 0 & 0 & 0 & 1  \\
 0 & 0 & 0 & 0 & 0 & 0 & 1 & 0  \\
\end{pmatrix}
\end{equation}
where we note the phase on state $|100\ra$. This unitary can be implemented in a depth of just 7 \cite{barenco1995}, a notable reduction. 
Still, such a ``pseudo-Toffoli'' must be used with caution \cite{barenco1995,DiVincenzo94_pstoff}. Though a mixer unitary built from such pseudo-Toffolis may yield the same sparsity pattern, the phases of some computational basis states would be flipped for every mixing step of QAOA. This would make the connection between QAOA and AQO quite tenuous, as phases would be significantly modified outside the subroutine that exponentiates the cost function. Further analytical and numerical investigations would be required to understand the effect of replacing Toffoli gates by $U_{\rm{pT}}$ or similar unitaries.


\section{Combinatorial problems}\label{sec:problems}


Here we consider five prototypical classes of discrete combinatorial problems as 
concrete examples of applying DQIR: graph coloring, the traveling salesperson problem, factory/machine scheduling, financial portfolio rebalancing, and integer linear programming. 
For each problem we show how its domain, feasible subspace, and cost Hamiltonian may be naturally represented 
with DQIR, directly enabling the application of various quantum algorithms including but not limited to those listed in Sec.~\ref{sec:qalgs}.
%
Numerical results concerning the 
resource requirements for each problem are presented in Section~\ref{sec:results}. There are of course numerous further important classes of discrete optimization problem that we do not 
address explicitly here, but may be straightforwardly implemented in DQIR, including, for example, other 
problems 
related to
graphs such as graph partitioning or 
edge coloring~\cite{hadfield19_qaoa},
and lattice problems such as the closest 
or shortest vector 
problems~\cite{joseph2020two}. We refer the interested reader to \cite[App. A]{hadfield19_qaoa} for a compendium of relevant problems and example mappings.  

The purpose of this section is to two-fold. First, these examples are pedagogically useful when designing quantum approaches to broader real-world industrial problems. For instance, the concepts used for our simple factory scheduling problem are applicable to much more complex scheduling problems. Our second aim is to demonstrate through examples that DQIR provides a representation 
for discrete problems that is 
more compact, interpretable, and portable  
than previous direct-to-qubit 
approaches. We remark that while qudit generalizations of the Pauli matrices exist for arbitrary $d$ (see e.g. \cite[App. C]{hadfield19_qaoa}), mixers derived from them may be significantly more expensive in given encoding than the approach of Sec.~\ref{sec:mixers}.




\subsection{Graph coloring problems}\label{sec:coloring}
A variety of computationally challenging optimization problems 
can be related to colorings of graphs. Here we consider the problem of assigning colors to vertices such that the number of edges with differently colored ends is maximized.


When considering $d$ colors, and reformulating the problem as a minimization, 
count the number of 
adjacent vertices with equal values. In DQIR 
this gives the cost Hamiltonian
%
\begin{equation}
H_C = \sum_{(\alpha,\beta)\in E} EQ(x_\alpha,x_\beta)
\end{equation}
%
where 
$x_{\gamma}$ denotes the color on node $\gamma$, $EQ$ is defined in equation \eqref{eq:EQ}, and the sum is taken over the graph edges. 

When some of the state space is invalid (for example if there are 3 colors and an SB encoding is used), one may use either single-variable mixers 
and/or add a variable domain penalty $F_{SS}$.
Alternatively, when one-hot encoding is employed, we note that the condition that each node is singly colored translates to constraints on the allowed Hamming weights such that many of the single-qubit Pauli-Z terms can be excluded from the cost Hamiltonian as they simply yield an overall 
constant \cite{hadfield19_qaoa}.
%

Cost Hamiltonians for a variety of related problems
may be similarly constructed~\cite{hadfield19_qaoa}. This includes both other problems 
over vertex colorings of a graphs but with different cost functions and hard constraints, for example problems related to proper colorings such as approximating the graph's chromatic number or size of the largest properly colorable induced subgraphs, as well as problems over different domains such as edge colorings. In general graph coloring problems have a close connection to scheduling, routing, and planning problems as we elaborate on below.

%
%
%





\subsection{Traveling salesperson problem}\label{sec:tsp}


The well-known
traveling salesman/salesperson problem (TSP) is to minimize the total distance travelled 
on a round trip visiting each of $M$ cities exactly once. In the formulation we adopt here, a valid state is a permutation $\tau$ of the $M$ cities. The distances between cities is given by a distance matrix $d(\a,\b)$ of size $M \times M$. The 
problem may be expressed 
as minimization over permutations $\tau \in S_M$ of the cost function
\begin{equation}
\sum_{a=0}^{M-1} d \left( \tau(a),\tau(a+1) \right),
\end{equation}
%
which for 
each permutation $\tau$ gives the corresponding 
total distance traveled, and where $M \equiv (M \mod M) = 0$ in order to include the final distance $d(\tau(M-1),\tau(0))$. 
At 
the DQIR level, the problem is encoded as a list of $M$ integers. %
The cost Hamiltonian in DQIR is
\begin{equation}\label{eq:tsp_cost}
H_C = \sum_{\a=0}^{M-1} \sum_{k=0}^{M-1} \sum_{l<k} d(k,l) \mP\supr{k}_\a \mP\supr{l}_{\a+1}
\end{equation}
%


Note that equation \eqref{eq:tsp_cost} consists of at most two-variable terms in DQIR. 
%
To ensure that one stays on the feasible space of permutations, one may use PPMs after preparating an initial state in a superposition of permutations \cite{bartschi20_permut}, or introduce the penalty $F_{perm}$.




\subsection{Machine scheduling}\label{sec:sched}

Machine scheduling (or job sequencing) \cite{brucker2004sched,venturelli15_jobshop} is another problem class with a 
domain that can be defined as a permutation of integers,
though its cost function is substantially different from TSP. 


Here we consider a simple single machine scheduling (SMS) problem. 
We are given a set of $M$ jobs with three properties: processing times $\{p_k\}$, deadlines $\{d_k\}$, and weights $\{w_k\}$. For job $k$, the start time is $s_k$. 
Our goal is to determine start times that minimize total weighted lateness $\sum_k w_k T_k$, where the lateness of each job is $T_k = (s_k+p_k-d_k)$. 
Note that one may also express the scheduling 
configurations as a list of start times (instead of a permutation of job IDs) \cite{hadfield19_qaoa}, or otherwise, though we do not consider such 
representations here.
Importantly, a number of more sophisticated but closely related 
scheduling problem variants are studied in the literature, including ones with additional hard problem constraints and multiple machines ~\cite{brucker2004sched}.

We choose to use lateness \cite{hariri1997lateness,brucker2004sched,cai2007lateness,akyol2008earlytardy} in our problem definition here in order to consider as simple a cost function as possible. However, many other cost functions may be of interest in real-world 
scheduling problems, 
for instance tardiness or earliness, defined respectively as $\max \{0,s_k + p_k - d_k\}$ and $\max \{0,d_k - s_k - p_k\}$ \cite{brucker2004sched}. Note that because lateness is simply tardiness minus earliness, 
it may be viewed a special case of the latter quantities. Though we do not explicitly consider tardiness and earliness in this work, we note here that the quantum phase operator may be extended to the $\max\{0,f(x)\}$ operation 
using ancilla qubits. 
First one would compute $f(x)$ into the ancilla register using subroutines of Section \ref{sec:primitives}, 
which may include standard arithmetic operations, 
before performing 
operators controlled on the ancilla register value being positive to implement phase kickback, 
and finally uncomputing the first step to disentangle the ancilla register. 
 Alternatively, one may consider distinct representations of the problem domain~\cite{hadfield19_qaoa}. 






For SMS problem it is useful to introduce \textit{value primitives} 
for the processing times
\begin{equation}
\m A_\a^{proc} = \sum_{k=0}^{M-1} p_k \mP\supr{k}_\a
\end{equation}
and the deadlines
\begin{equation}
\m A_\a^{dead} = \sum_{k=0}^{M-1} d_k \mP\supr{k}_\a.
\end{equation}

In DQIR, a precursor for the start time operator can be written as an expression of value primitives and an indicator primitive,
\begin{equation}
s_{k,\alpha} = \m P\supr{k}_\a \sum_{\beta<\alpha} \m A_\b^{proc}
\end{equation}
where this operator's support on a classical state is non-zero only when the $k$th job is in the $\alpha$th position. 
This is a prototypical example of a value primitive being useful in mapping classical problems to the quantum representation of DQIR.

A full start time operator for job $k$ may be expressed as
\begin{equation}
s_k = \sum_{\a}^{M} s_{k,\a} = \sum_{\a}^{M} \m P\supr{k}_\a \sum_{\beta<\alpha} \m A_\b^{proc}
\end{equation}
%
%
Combining the above formulas, one way to compactly describe the cost function of total weighted tardiness is 
%
\begin{equation}\label{eq:sms}
\begin{split}
H_C &= \sum_{k} s_k + \sum_\a^{M} \left( \m A_\a^{proc} - \m A_\a^{dead} \right) \\
&= \sum_{k} \left( \sum_\a^M \m P\supr{k}_\a \sum_{\beta<\alpha} \m A_\b^{proc} \right) \\&+ \sum_\a^{M} \left( \m A_\a^{proc} - \m A_\a^{dead} \right).
\end{split}
\end{equation}
%





We highlight again the compactness of this expression, that it is encoding-independent, and that it written entirely in terms of the primitives of DQIR. 
Because SMS is a permutation problem, mixers, penalties, and/or initial states need to be introduced as appropriate, analogously to the case of TSP.

Note that when Gray or SB is used, $\sum_k \m P\supr{k}_\a$ in equation \eqref{eq:sms} is equal to the identity in qubit space if $\ceil{\log_2{d}}=\log_2{d}$. Hence when a compact code is $d$ is a power of 2, this will lead to very favorable cancellations only in these compact encodings, as we will see in the numerical results.





\subsection{Financial portfolio rebalancing}\label{sec:finance}


In computational finance, the task of portfolio rebalancing is an optimization problem. Here we use a simple model previously proposed in the context of QAOA \cite{hodson19_portf}. For a set of financial stocks separated into ``lots'' of discrete quantities, and given the previous portfolio position and a set of parameters, the goal is to determine whether to have a long, short, or no-hold position for each lot. The relevant parameters are the solution vector $\mathbf  z \in \{-1,0,+1\}^M$, the normalized risk-return function $C_{RR}(\mathbf z)$, the normalized trading cost function $C_{RC}(\mathbf z)$, the number of lots $M$, the number of lots one may invest in $D$, a risk parameter $0 \leq \lambda \leq 1$, the normalized asset returns covariance matrix $\sigma$, the normalized average asset returns vector $\mu$, the previous portfolio position $\mathbf  y \in \{-1,0,+1\}^M$, and the normalized cost $T$ of trading. 

The optimization function is
\begin{equation}
H_C = C_{RR}(\vec z) + C_{TC}(\vec z)
\end{equation}
where
\begin{equation}
C_{RR}(\textbf{z}) = \lambda \sum_\alpha^M \sum_\beta^M \sigma_{\alpha\beta} z_\alpha z_\beta - (1 - \lambda) \sum_\alpha^M \mu_\alpha z_\alpha
\end{equation}
and
\begin{equation}
C_{TC}(\textbf{z}) = \sum_\a (1 - \delta(z_\a-y_\a))T,
\end{equation}
where $\delta()$ is the Dirac delta function. We now write down DQIR expressions for the classical $C_{RR}$ and $C_{TC}$.
An arbitrary choice for labeling the three possible states is to use \{long, no-hold, short\} $\mapsto \{|0\ra,|1\ra,|2\ra\}$, which can be expressed via the value primitive
\begin{equation}
\begin{split}
\mA_\a^z &= \sum_i z_i |i \kb i|_\a \\
&= (-1)  |0 \kb 0| + (0) |1 \kb 1 | + (+1) |2 \kb 2| \\
&= |2\ra\la2| - |0 \ra\la 0|
\end{split}
\end{equation}
%







In DQIR the neat expressions are
\begin{equation}
C_{RR} = \lambda \sum_\alpha \sum_\beta \sigma_{\alpha\beta} \mA_\alpha \mA_\beta - (1 - \lambda) \sum_\alpha \mu_\alpha \mA_\alpha
\end{equation}
and
\begin{equation}
\begin{split}
C_{TC} = \sum_\a^M \left(1- \m P\supr{y}_\a \right)T
\end{split}
\end{equation}
where we note that any delta function between a constant and a variable may be replaced by $\delta(j_0-z)_\a \mapsto \m P\supr{j_0}_\a$. 
Finally, we have the constraint
\begin{equation}
\sum_\alpha z_\alpha = D
\end{equation}
which may be treated as a hard constraint by choosing appropriate initial state and mixers, or as a soft constraint by introducing a $F_{sum}$ penalty (equation \eqref{eq:f_sum}). 





\subsection{Integer programming}\label{sec:ilp}


Because many 
industrially relevant problems tend to be cast as integer programming and related problems \cite{ilp_book,leipold2021driver}, 
%
here we outline the canonical form of ILP and briefly summarize its implementation in DQIR. Given a rectangular matrix $A$ and constant vectors $\mathbf  b$ and $\mathbf  c$, the problem is to \textit{maximize} $\mathbf  c \cdot \mathbf  x$ under the two constraints 
\begin{equation}
A \mathbf  x \leq \mathbf  b
\end{equation}
and $\mathbf  x \geq \mathbf  0$. 
The cost function for canonical ILP is simply represented as
\begin{equation}
H_C = \sum_\alpha c_\alpha \m A_\a^{\rm{ILP}}
\end{equation}
where the value primitive is the previously defined number operator $\m N = \sum_k^d k \m P\supr{k}$. 
Considering that this cost function is ``one-local'' in DQIR space, most of the algorithmic complexity appears via the problem's constraints, \textit{e.g.} via introduction of constructs such as $F_{\text{lin}}$ in equation \eqref{eq:f_ilp_i}. 

We note that it may be computationally hard in general to even produce an initial state that contains the full feasibly subspace while excluding the infeasible subspace. Therefore, it appears likely that penalties (Section \ref{sec:penal}) will be necessary in NISQ implementations of ILP that use the approach of this subsection, else other approaches may be applicable only in limited cases. While we do not expect quantum computers to efficiently solve NP-hard problems such as ILP it may prove worthwhile to explore problem variants or restricted settings where quantum approaches may become viable.

\section{Numerical Study of Encoding Choice}\label{sec:results}




The purpose of this section is to perform a preliminary but extensive numerical investigation of the differences in behavior between the different encodings in terms of quantum resources (specifically circuit depths), with a focus on the subroutines in Sections \ref{sec:algo_compon} and \ref{sec:gdm}, 
and applied instances of 
problems considered in Section \ref{sec:problems}. The observations and results of this section may help guide practitioners in deciding which encoding to use when designing suitable algorithm implementations for particular quantum devices. 
We emphasize that, as 
illustrated in Figure \ref{fig:pareto}, different encodings require significantly different quantities of qubits---hence the shortest-depth encoding 
may not always be the encoding of choice for a given set of hardware. Note that we assume all-to-all connectivity in the results presented here.
%
%
%


With the exception of the strict GDPM mixers, this preliminary resource analysis involves compiling 
the matrix exponential $\exp(-i\beta H)$ for a Hermitian operator $H$ and real constant $\b$. The general connection to quantum algorithms is that the standard formulations of AQO and QAOA are directly implemented in terms of such operator exponentials; hence these numerical results may aid in making preliminary encodings choices when implementing AQO or QAOA. 
For QAOA, as discussed a variety of different mixers are possible with different resource tradeoffs beyond those we consider here. 
The connections to QITE and non-QAOA ansatzes are less obvious and thus our results are less applicable to those two classes of algorithm.

The operators for which we calculate circuit depths should be placed into three distinct categories: (a) exact exponentials of diagonal functions, (b) approximate mixers as implemented in equation \ref{eq:trottermixer}, and (c) strict mixers. The former two operator categories use the same compilation procedure, which is described in Section \ref{sec:sw}: after a Hermitian operator is converted to the Pauli representation generated by a given encoding, each Pauli term is exponentiated as the simple first-order product formula of equation \ref{eq:trottermixer} with an ordering choice described below. This product formula leads to exact exponentials only in category (a) because all the terms are diagonal and thus commute. 
On the other hand, the calculated depths for the strict mixers of category (c) are determined as described in Section \ref{sec:decomp}. 
Results in Sections \ref{sec:simple} and \ref{sec:HC_num} belong to category (a), while those in Section \ref{sec:mix_numer} belong to categories (b) and (c). We consider DQIR variable sizes $2 < d \leq 16$, from which some preliminary trends emerge.
We plot the resulting circuit depths for the different encodings and discuss the observed relative advantages.


We note that there are many Hamiltonian simulation algorithms for implementing the exponential of a Hermitian operator, each with various resource and accuracy trade-offs \cite{suzuki76,berry15_blackbox, berry15_hamsim_qwalk,Berry2015, qsp2017,  low2019_qubitization, Childs2019_random, Campbell2019_random, childs19_theoryoftrotter}. In this work, in an attempt to provide some preliminary guidance specifically for nearer-term quantum optimization, we chose a recently proposed Hamiltonian simulation technique \cite{schmitz2021_popr} that is designed to produce short-depth circuits for early generations of quantum hardware. The trends observed in our numerical results may differ when using other Hamiltonian simulation algorithms, including methods that require advanced fault-tolerant quantum hardware.



\subsection{Compiling product formulas from DQIR}\label{sec:sw}




We implemented a prototype of DQIR in Python, which we used to produce our numerical results. The code is built on three object classes: one to represent a discrete variable, one for defining the multivariate space, and one for describing and manipulating an arbitrary operator as a sum of products of local primitives. The latter class is based on the use of a Python dictionary, where keys are operator strings and values are coefficients which may be numeric or symbolic. High-level functions return a DQIR object for each combinatorial problem type and auxiliary operators such as mixer generators. Algebra routines are built into the classes, allowing for multiplication and addition of any operators as well as replacing symbolic coefficients with numerical values. We use subroutines from mat2qubit \cite{mat2qubit}, Scipy \cite{scipy}, OpenFermion \cite{openfermion}, and SymPy \cite{sympy}. The examples included in the mat2qubit package give explicit code for producing these Hamiltonians, and we also include the qubit-encoded Hamiltonians and raw quantum circuits in the Supplemental Materials. Terms are cancelled and combined inside DQIR, which reduces the amount of computational algebra that need be performed in the final Pauli representation; this is especially important when one's goal is to compare many encodings. We implemented functionality for returning a complex operator's full matrix representation, allowing for the study of matrix properties.

For qubits, converting to the Pauli representation is automated with the help of the Intel Quantum SDK \cite{khalate2022qsdk}. After each discrete variable is assigned an encoding, the DQIR-to-qubit encoding procedure of Section \ref{sec:dqir2qub}
is used. 
This yields a weighted sum of terms $\sum_v G_{M,v}$ as in equation \eqref{eq:gen_decomp}, where each $G_{M,v}$ is a Pauli string with a real coefficient. Our task is to compile a circuit that performs either an exact (for cost functions) or approximate (for mixers) 
simulation of the exponential $\exp({ \sum_v-i \beta G_{M,v}})$. The circuit depths that we report are for the simple product formula $\prod_v e^{ -i \beta G_{M,v} }$ for arbitrary real $\beta$, where the ordering of terms is determined using an algorithm that attempts to minimize circuit depth (discussed below). For all diagonal operators (cost functions, penalties, and diagonal primitives), this procedure provides \textit{exact} exponentials, because all $G_{M,v}$ commute with each other. For non-diagonal mixer Hamiltonians this procedure leads to an \textit{approximate} mixer that may produce leakage as described in Section \ref{sec:inexact-trott}. We do not necessarily recommend the use of these instances of approximate mixer, but perform this analysis in order to have a point of comparison against our novel GDPMs). The circuit depths we present are independent of the scalar $\beta$ in the exponential.  
Note that circuit depths for the strict mixers are based on Section \ref{sec:decomp}, \textit{not} based on the compilation procedure of this section.


From the Pauli representation we produce first-order product formulas using an algorithm \cite{schmitz2021_popr}
based on Pauli Frame Graphs (PFG). 
In this approach, one does not ``uncompute'' the change of basis for each Pauli term, but instead moves directly to another Pauli frame in order to exponentiate the next term. Optimizing the circuit depth 
can then be recast in terms of a graph search problem. This PFG-based approach has been shown to produce significantly shorter circuits for a variety of physics and chemistry Hamiltonian classes, as compared to more standard circuit construction methods. 

It is worth noting that the use of a gate set that natively performs the continuous two-gate operation $\exp(-i \theta ZZ)$ \cite{ryan08nmr} may lead to shorter depths as well as circuit times that are more directly dependent on the magnitude of the coefficients. However, in our view such a gate set is unlikely to be the one that is used in most commercialized quantum hardware \cite{versluis2017scalable,lekitsch2017blueprint}. While some platforms are capable of arbitrary-angle multi-qubit entangling gates, and such operations can vary in time of operation, ``continuous'' control is antagonistic to arbitrary operation time due to the difficulties of system-wide clock synchronization. Thus, in practice, operations ought to be fixed to integer multiples of the system clock, where idle time is typically inserted to make up the difference. Thus at best, small angles will still likely be limited to the time of a single clock cycle. 



When reporting results below, our calculated circuit depths result from the PFG decompositions, where our code uses a gate set of arbitrary one-qubit rotations and the 9 entangling gates defined in reference \cite{schmitz2021_popr}, which includes the CNOT gate. 
(Converting these entangling gates to CNOT can be done using only one-qubit gates; because such one-qubit gates may be fused with existing adjacent one-qubit gates, the reported circuit depths are similar 
to using the gate set of CNOT and arbitrary one-qubit rotations.) Single-variable strict mixers were designed by Algorithm \ref{algo:sgl-var-mix}, and the resulting unitaries were decomposed based either on the PFG algorithm or on the circuit decompositions summarized in Section \ref{sec:decomp}, whichever yielded shorter depth.




\subsection{Primitives and penalties}\label{sec:simple}

\begin{figure*}[t]
    \centering
    \includegraphics[width=.8\textwidth]{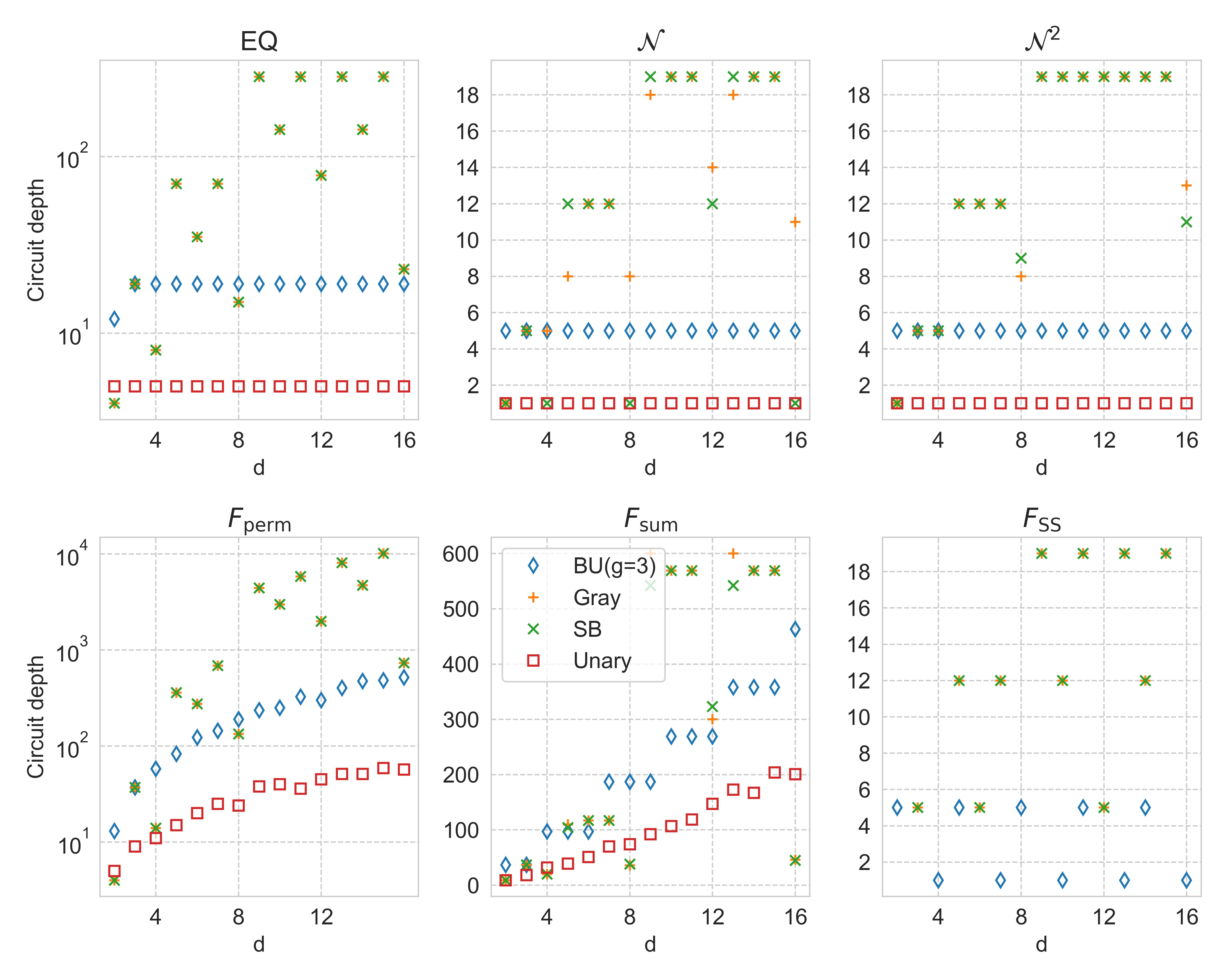}
    \caption{Circuit depths for exponentiating various 
    diagonal primitives and penalties with varying $d$. As all terms commute in the diagonal cost functions, these depths correspond to exact exponentials.}
    \label{fig:simple_funcs}
\end{figure*}



Here we present numerical results for diagonal operators that are used as building blocks in the construction of cost functions, \textit{i.e.} we study primitives from Section \ref{sec:primitives} and penalties from Section \ref{sec:penal}. Figure \ref{fig:simple_funcs} shows circuit depth plotted against variable size, for diagonal operator building blocks $EQ$, $\m A_n$, $\m A_n^2$, $F_{perm}$, $F_{sum}$, and $F_{SS}$ of Section~\ref{sec:dqir}. We remind the reader that unary requires $d$ qubits, compact codes require $\ceil{\log_2 d}$ qubits, and block unary interpolates between them (see Figure \ref{fig:pareto}). Thus the shortest-depth circuit is not necessarily the most appropriate choice if a given quantum device is space-limited. Results here assume all-to-all connectivity; circuit depths will be different when considering a different hardware topology \cite{sawaya20_connect}, an important 
aspect beyond the scope of this work.

For most diagonal operators considered here, unary provides the shortest-depth circuits. However, there are many important exceptions apparent for $\m A_n$ 
and $F_{\text{sum}}$ 
These exceptions are notable for the following reason. When the circuit depths for the compact codes are equal to or less than the depths for unary, this means a compact code is likely to be superior on most hardware, as it would require \textit{both} shorter depth and fewer qubits. However, the circuit depth of the \textit{full} quantum algorithm (not just one of its subroutines) must be analyzed to determine whether this is the case---we do not perform such an analysis of full quantum algorithms in this work, but we note that in the context of Hamiltonian simulation there are some problem instances for which compact codes out-performed unary in terms of both depth and space \cite{sawaya20_dlev}.

Because of favorable term cancellation in qubit space, compact codes tend to be most useful when $d$ is a power of 2, as discussed in previous work \cite{sawaya20_dlev}. This trend is exhibited in the examples of Figure \ref{fig:simple_funcs}, where compact codes are more likely to be competitive with other encodings when $\log_2 d$ is an integer.

Though the Gray and SB encodings often yield the same circuit depth, in some cases they have different depths, for example in $\m N$ and $F_\text{sum}$. This highlights the fact that in some instances it may be worth considering multiple compact encodings. 
A practitioner may choose to explore more of the (exponentially many) compact encodings before choosing the shortest-depth one.

The reason for BU's intermediate depth is that the locality of the diagonal operators is smaller than the compact code but larger than unary. For a given quantum device, block unary is an optimal choice only when two necessary conditions are met. First, the depth for BU must be less than that of the compact codes. This is true for most $d$ values in the simple diagonal functions considered here. Second, the hypothetical device must be qubit-limited, such that there are not enough qubits available to use a unary code. Because our results show that (often) BU is intermediate in both depth and in qubit count as compared with the other encodings, there is a strong chance that BU will be the optimal choice for \textit{some} near-term hardware parameters (those with moderate depth and moderate qubit counts).

The three types of penalties considered show similar behavior; namely, unary is usually shortest-depth and it is common for BU to have shorter depth than compact codes. Note that the single-variable validity penalty $F_{\text{SS}}$ is the \textit{null} operator in the following cases: in compact codes when $d$ is a power of 2, in block unary when $d$ is divisible by $g$, and always for the unary encoding because we are assuming that validity-preserving operators are used for unary algorithms. Both the permutation and summation penalties involve a quadratic number of two-variable terms, which is the reason for their larger circuit depths. Note that the depth of the permutation penalty is strongly dependent on the number of discrete variables $M$ (\textit{e.g.} number of cities). This suggests that problems for which $M$ is not a power of 2 ought to be reformulated---this could be accomplished in the TSP by adding cities with very small distance to an existing city, and in scheduling this could be accomplished by adding extra tasks with very small $p_k$.

Finally, we highlight one clear example where \textit{penalty exchanging} (Section \ref{sec:pen_exch}) would reduce overall circuit depth. When $d=7$ with compact codes, the $EQ$ operator yields a depth of 70 while single-variable validity penalty $F_{SS}$ yields depth 12. Hence one can reduce depth by implementing the $d=8$ version of $EQ$ (depth 15) while introducing a validity penalty for $d=7$, which results in a net overall depth reduction. As mentioned, shorter depth does not always lead to improved algorithms---any modification of a subroutine ought to be tested for performance in a real algorithm.




\subsection{Full cost functions} \label{sec:HC_num}

\begin{figure*}[t]
    \centering
    \includegraphics[width=0.8\textwidth]{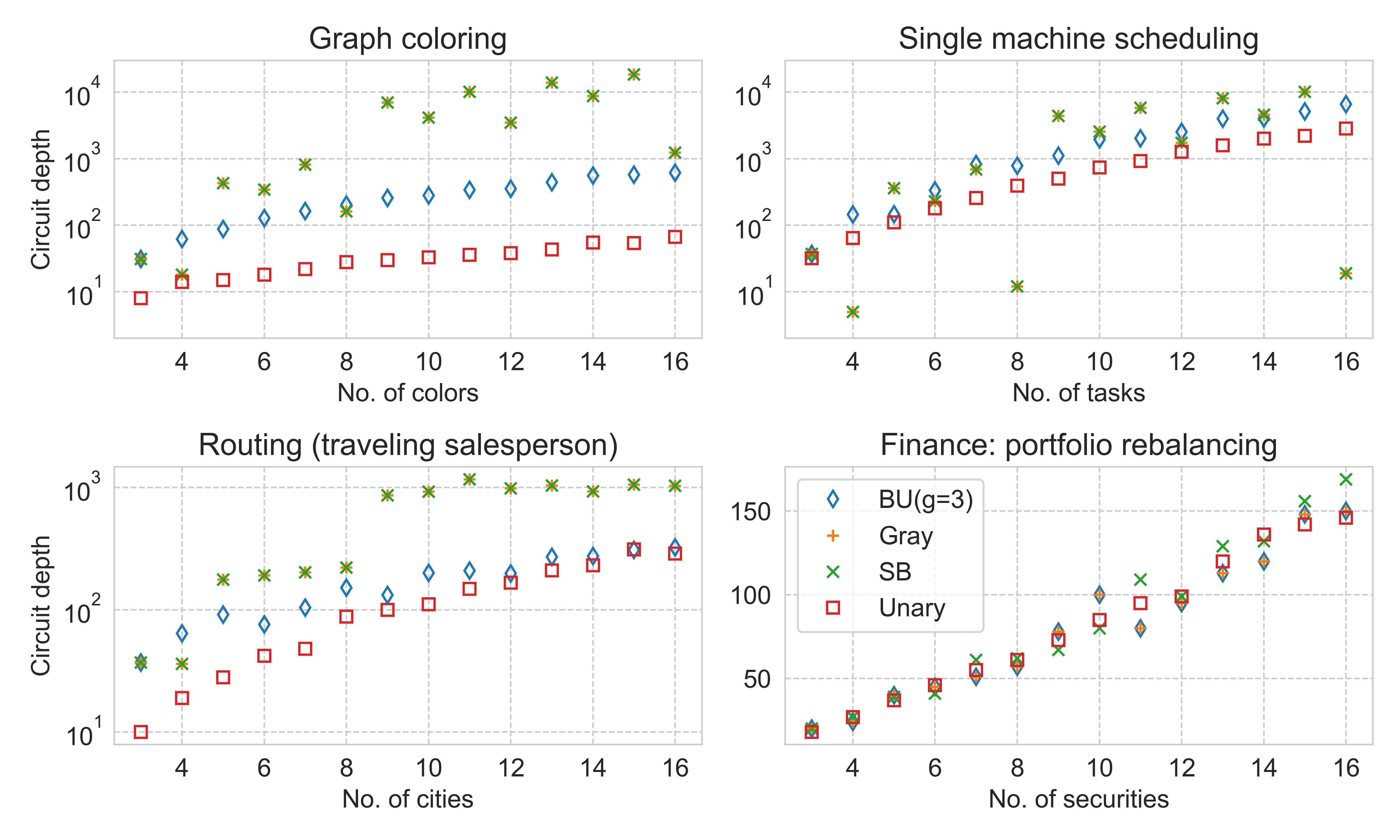}
    \caption{Circuit depths 
    for implementing the exponential of four cost Hamiltonians. As all terms commute in the cost functions, these depths correspond to exact exponentials. The graph coloring Hamiltonians use complete graphs with $d$ nodes. All portfolio rebalancing problem instances have $d=3$.}
    \label{fig:cost_hams}
\end{figure*}

Here we consider circuit depths for implementing the exponential of a cost function, a subroutine that is necessary for executing AQO and QAOA (summarized in Section \ref{sec:qalgs}). Figure \ref{fig:cost_hams} shows circuit depths for exponentiating four classes of cost functions defined in Section \ref{sec:problems}: graph coloring, scheduling, routing, and portfolio rebalancing. The problem parameters used are defined in the text below. Though these problem instances should be considered as toy models, the numerical results are useful for understanding trends in $d$ and in encoding choice that may also be applicable to more realistic real-world problems. 
%
%

Our graph coloring problem instances use complete (\textit{i.e.} fully connected) graphs for which the number of nodes equals the number of colors $d$, \textit{i.e.} $d$ scales linearly with the 
number of variables $M$. 
We use this class of graphs as a \textit{worst-case} example of scaling behavior. Resource requirements will be substantially different for example when $d$ is constant and independent of $M$. In real-world problems, the circuit depth would be heavily dependent on both the class of graph (\textit{e.g.} complete, regular, multipartite, \textit{etc.}) and how the number of colors $d$ scales with $M$. Graph coloring is ``two-local'' in the sense that all terms are products of at most two DQIR variables, and in the case of complete graphs the number of terms scales quadratically. The results, which follow the trends of the $EQ$ operator, show that while unary and BU encodings follow clean monotoic trends, in compact codes the depth shows a much less consistent trend. For example, for SB and Gray there is a large drop in depth from $d=7$ to $d=8$. Expanding the domain and using penalty exchanging may be appropriate in such cases.

Machine scheduling (or factory scheduling) is one of two problems we present where the solution must be a permutation of integers. The substantial depths observed here result from the quadratic number of two-variable terms in the cost Hamiltonian. At $d=4,8,16$, the depths for compact (SB and Gray) encodings are over an order of magnitude lower than the unary depths. This is because, when $d$ is a power of 2 and a compact encoding is used, the $\sum_k \m P\supr{k}_\a$ term of equation \eqref{eq:sms} \textit{in qubit space} is equal to the identity, which leads to massive simplifications and term cancellations in the Hamiltonian. We hasten to note that this does \textit{not} necessarily imply that the compact codes are superior to unary for machine scheduling. An algorithm implementation requires not just the cost function but also either a $F_{perm}$ penalty or PPMs, the versions of which have presented are much higher depth for compact than for unary (see Figures \ref{fig:ppm_circuits} and \ref{fig:simple_funcs}).

However, we have not extensively explored the design of PPMs in this work. We are hopeful that a future design of PPMs that operate on fewer qubits (see the hypothetical case on the left of Figure \ref{fig:hard-ppm-mixer}) may lead to PPMs that are much shorter-depth when decomposed. Though their existence is only speculated, if such shorter-depth PPMs are possible, it may result in the compact code being both the lowest-space and lowest-depth choice in QAOA for machine scheduling.

TSP (routing) is also a permutation problem composed of two-variable terms and all-to-all connectivity. As in the case of machine scheduling, higher-depth penalties and/or mixers are unavoidable when solving the problem, because one must stay in the permutation space. The numerical trend matches what was expected, with circuit depths highest for compact and lowest for unary. This is another case where the results appear to suggest that BU has a strong chance of being viable, in cases where there are not enough qubits available for the unary encoding.

In the financial portfolio rebalancing problem, $d=3$ regardless of the number of variables. Therefore BU$^{\rm{Gray}}_{g=3}$ gives identical results to the Gray code, as they both use 2 qubits per variable. 
The circuit depths for all encodings are very similar, which leads us to conclude that the more space-efficient compact codes are preferred for this Hamiltonian class, especially considering the short-depth $d=3$ compact mixers that we present. Note that the Gray code is similar to the encoding used in the original study of QAOA for portfolio rebalancing \cite{hodson19_portf}.

Notably, as observed in the last section, these full cost functions (excluding portfolio rebalancing) show that that there is a potential role for block unary---certainly more so than previous studies of physics Hamiltonians would have suggested \cite{sawaya20_dlev}. This is because when the depth of BU lies in between compact and unary, there exists some set of hardware parameters for which block unary is the optimal choice. As previously stated, BU is the optimal choice when there are not enough qubits to use unary, but there are enough qubits to implement BU or compact codes.


\subsection{Mixers}\label{sec:mix_numer}

\begin{figure*}[t]
    \centering
    \includegraphics[width=.8\textwidth]{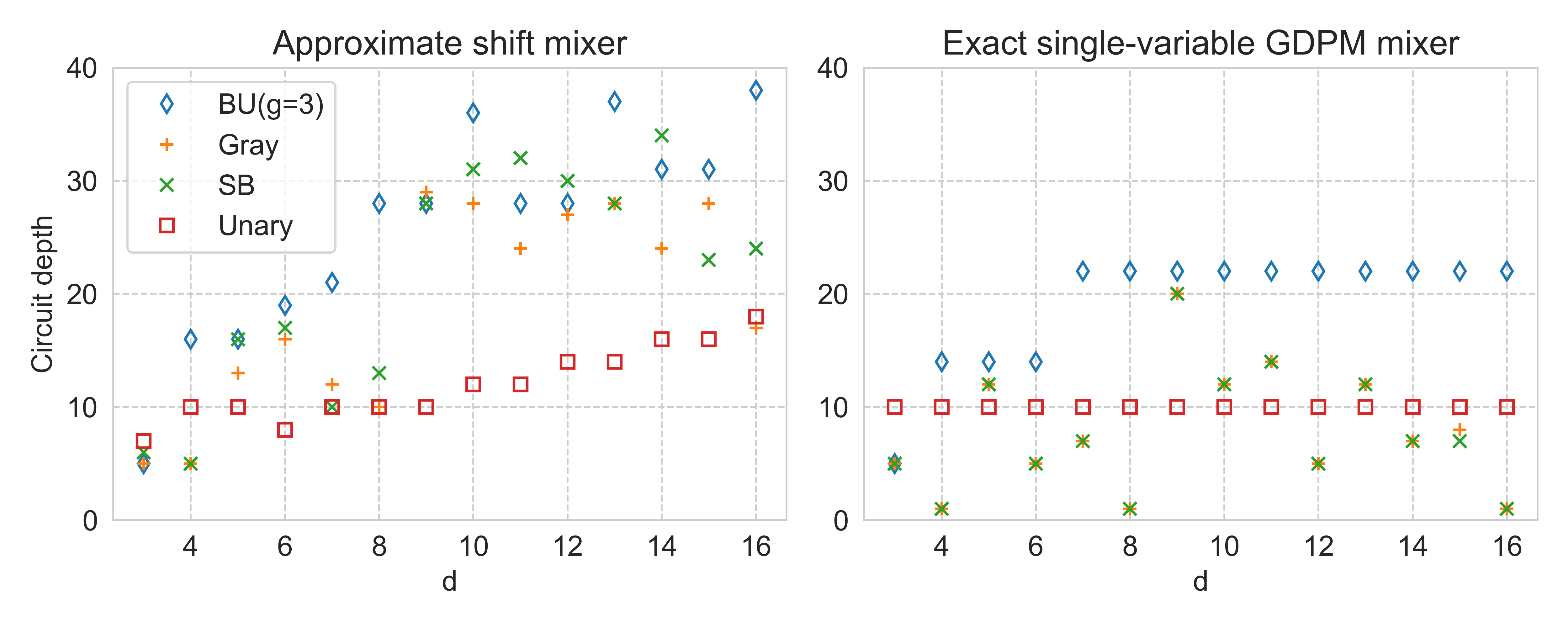}
    \caption{ \textbf{Left}: Circuit depths for approximate single-variable mixers (which cause some leakage) defined as the exponential of equation \eqref{eq:shift_ham}. \textbf{Right}: Strict GDPM single-variable mixers designed using Algorithm \ref{algo:sgl-var-mix}, decomposed into one- and two-qubit gates. }
    \label{fig:mix_err}
\end{figure*}

Figure \ref{fig:mix_err} shows circuit depths for implementing approximate (\textit{i.e.} leaky) shift mixers (left) and strict (leakage-free) single-variable GDPMs (right) designed by Algorithm \ref{algo:sgl-var-mix}. The approximate shift mixers are simply the first-order product formula that implements  decomposition of the exponential of equation \eqref{eq:shift_ham}, with the term ordering and depths determined using the previously mentioned PFG approach \cite{schmitz2021_popr}. In the exact mixers, depths are determined by whichever decompositions outlined in Section \ref{sec:decomp} produce the lowest depths.

In nearly all cases, the naive approach of approximating the exponential of a mixer Hamiltonian yields larger depths than our designed exact mixers. This means that the single-variable GDPM designs are superior \textit{both} in terms of depth and in terms of not producing leakage. This suggests that our algorithmically designed single-variable mixers will find utility in real algorithms, though we leave numerical analysis of these mixers' performance in a QAOA simulation to future work.

BU mixers were designed by using two-qubit mixers within each block, and connecting each two-qubit block with doubly-controlled $A_\phi$ gates. The latter four-qubit gates connect two blocks by mixing adjacent cross-block integers. For example, states $|2\ra \mapsto |0010\ra$ and $|3\ra \mapsto |0100\ra$ are mixed using such a four-qubit gate. The depths for this four-qubit gate were calculated using the PFG method. The resulting depths are largest for BU in all cases $d>3$, but the discrepancy is not large enough to overwhelm the differences in cost function depths shown in the previous two sections. These are not necessarily the optimal mixer designs for BU.

In compact codes, the figure shows the mixer depth is equal to one when $d$ is a power of 2 ($d=4,8,16$). This is because the simple mixer of equation \eqref{eq:sbm} is used. This short depth, along with the fact that circuits for cost function exponentials also tend to be shorter depth at these same $d$ values (Figure \ref{fig:cost_hams}), shows that SB and Gray may sometimes be more competitive with unary than expected, while requiring far fewer qubits. Finally, we highlight that the GDPM compact (Gray and SB) mixers are even shorter depth than unary mixers for the majority of $d$ values. 

\section{Conclusions}\label{sec:concl}



In this work we have 
constructed and analyzed a large set of quantum subroutines relevant to solving optimization problems defined on domains of discrete (\textit{e.g.} integer) variables. We introduced an intermediate representation (DQIR) which facilitates the synthesis of a variety of algorithm components into qubit-based hardware, though these procedures may be used for qu\textit{d}it-based hardware as well. Previous work had analytically derived operators for each encoding, whereas our method automates the implementation of any arbitrary integer-to-bit mapping by building on previous techniques \cite{hadfield19_qaoa,sawaya20_dlev}.

The first advantage of our approach is that it provides a compact, flexible and readily interpretable representation of discrete optimization problems; one may fully define discrete problems and algorithmic components 
before considering any hardware implementation. Second, one may automate the process of ``screening'' any number of potential hardware mappings, 
comparing circuit depths or other measures of resource efficiency for 
different encodings. 
In addition, 
we have provided numerical and conceptual guidance for which operator components are most useful for which algorithmic approaches to a given problem class. We suggested several best practices regarding the interplay between initial states, preservation of feasibility and validity, and encoding choice.

Our numerical results yielded several rules of thumb regarding encoding choice. First, though the unary (one-hot) encoding is often the lowest-depth and highest-space choice, there are some subroutines for which the compact codes (\textit{e.g.} Gray and standard binary) are 
advantageous both in terms of qubit counts and circuit depth. Second, the block unary code, because its depth usually lies between the depths of compact and unary encodings, indeed often sits on the space-depth Pareto front (Figure \ref{fig:pareto}). This means that BU would in fact be the optimal choice for combinatorial problems for some hypothetical hardware parameters; this was somewhat surprising, as BU appears less likely to be useful in physics simulation \cite{sawaya20_dlev}. Third, the choice of bitstring for labeling each discrete value in compact codes can affect circuit depths---this is shown by the differences in depth sometimes observed between Gray and standard binary. Fourth, compact codes are much shorter depth when the variable domain $d$ is a power of 2. Finally, our numeric results highlighted that penalty exchanging, in which the the variable domain size is modified in exchange for introducing a penalty, may be an important strategy in algorithm design.

Separately, we have introduced criteria and approaches for designing a new class of QAOA mixers, which we call graph-derived partial mixers (GDPMs). Our computationally designed single-variable GDPM mixers provide systematic low-depth construction of exact mixers 
that preserve a desired feasible subspace. The very short depths of GDPMs for compact codes, even when $d$ is not a power of 2, lead us to be optimistic about the use of compact codes for quantum optimization.

There are several important open directions for future study. While we have implicitly assumed all-to-all connectivity throughout, the constructions of this work may be further tailored to a given hardware topology. Similarly, the generalization to quantum error correcting codes may be straightforward in some cases. In either case, 
the specific details may affect which encoding is preferable. In particular, the type of basic quantum gate which is considered "expensive" may be quite different between the near-term and fault-tolerant settings.
Using DQIR, automated compilation of many encodings, and the low-level quantum subroutines introduced in this work, some of these questions can begin to be 
addressed systematically.

In terms of algorithms, performance is ultimately the most tantalizing question, which has so far proven challenging to analyze. As improved quantum hardware becomes available better, empirical evidence should help further guide algorithm design and choice. For QAOA, many questions remain as to how to select the best mixer for a given problem with fixed quantum resources. For example, while alternative mixers requiring deeper circuits than the ones considered in Sec. 4 may in some sense provide more efficient mixing, their cost may limit the number of implementable QAOA layers; for comparable circuit depths it is not clear which approach will ultimately result in better performance.


\section*{Acknowledgements}\label{sec:ackn}
This research used the resources of the National Energy Research Scientific Computing Center (NERSC), a U.S. Department of Energy Office of Science User Facility located at Lawrence Berkeley National Laboratory, operated under Contract No. DE-AC02-05CH11231 using NERSC award DDR-ERCAP0018781. SH is grateful for support from the NASA Ames Research Center, 
under NASA Academic Mission Services (NAMS) Contract no. NNA16BD14C, from the DARPA ONISQ program under interagency agreement IAA 8839 Annex 114, and from the DARPA RQMLS program under interagency agreement IAA 8839 Annex 128.

\bibliographystyle{unsrt}

\bibliography{refs}

\end{document}